\documentclass[12pt]{article}

\usepackage{youngtab}

\usepackage{cite}
\usepackage{float}
\usepackage{amsfonts}
\usepackage{amsmath}
\usepackage{amsbsy}
\usepackage{graphicx}
\usepackage{amssymb}
\usepackage{amsthm}
\usepackage{bm}
\usepackage{epsfig}
\usepackage{latexsym}
\usepackage{pdflscape}
\usepackage{color}
\usepackage{here}
\usepackage{graphicx}
\numberwithin{equation}{section}

\allowdisplaybreaks

\Yboxdim{5pt}

\DeclareMathOperator{\tr}{tr}
\DeclareMathOperator{\Str}{Str}
\DeclareMathOperator{\Det}{Det}
\DeclareMathOperator{\im}{Im}
\DeclareMathOperator{\diag}{diag}

\newcommand{\W}{{\Gamma{}}}
\newcommand{\llangle}{\langle\!\langle}
\newcommand{\rrangle}{\rangle\!\rangle}

\newcounter{aff}

\begin{document}
\begin{titlepage}
\begin{flushright}
{\footnotesize OCU-PHYS 476, YITP-18-14}
\end{flushright}
\begin{center}
{\LARGE\bf 
Two-Point Functions in ABJM Matrix Model
}\\
\bigskip\bigskip
{\large
Naotaka Kubo\,\footnote{\tt naotaka.kubo@yukawa.kyoto-u.ac.jp}
\quad
and
\quad
Sanefumi Moriyama\,\footnote{\tt moriyama@sci.osaka-cu.ac.jp}
}\\
\bigskip
${}^{*}$\,{\it Center for Gravitational Physics, 
Yukawa Institute for Theoretical Physics,}\\
{\it Kyoto University, Sakyo-ku, Kyoto 606-8502, Japan}\\[3pt]
${}^{\dagger}$\,{\it Department of Physics, Graduate School of Science,}\\
{\it Osaka City University, Sumiyoshi-ku, Osaka 558-8585, Japan}\\[3pt]
${}^\dagger$\,{\it Osaka City University Advanced Mathematical Institute (OCAMI),}\\
{\it Osaka City University, Sumiyoshi-ku, Osaka 558-8585, Japan}
\end{center}

\begin{abstract}
We introduce non-trivial two-point functions of the super Schur polynomials in the ABJM matrix model and study their exact values with the Fermi gas formalism.
We find that, although defined non-trivially, these two-point functions enjoy two simple relations with the one-point functions.
One of them is associated with the Littlewood-Richardson rule, while the other is more novel.
With plenty of data, we also revisit the one-point functions and study how the diagonal BPS indices are split asymmetrically by the degree difference.

\centering\includegraphics[scale=0.6,angle=-90]{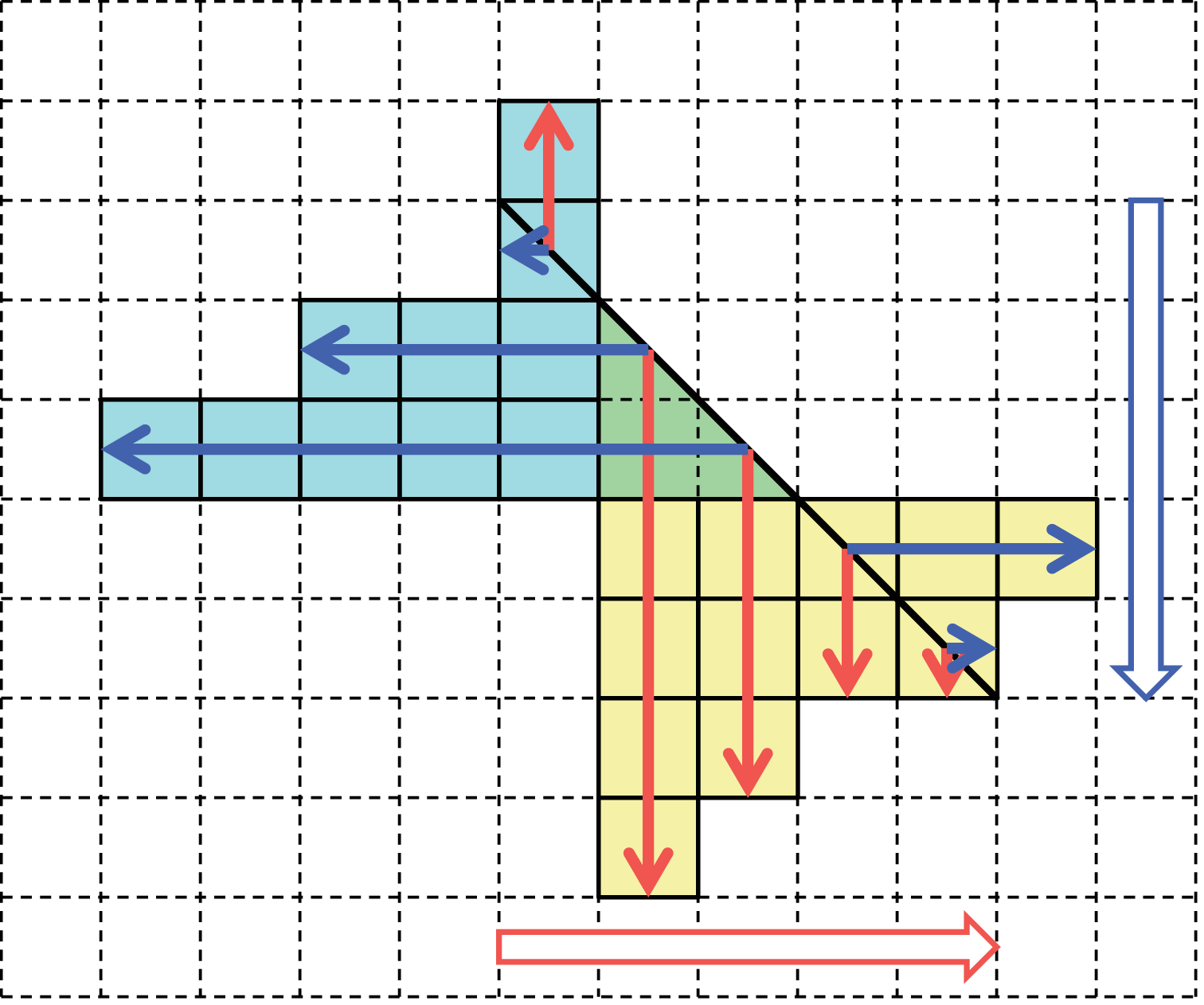}
\end{abstract}

\end{titlepage}
\setcounter{footnote}{0}

\tableofcontents

\section{Introduction}

Recently, there is a great progress in the study of the correlation functions on the M2-branes.
The most important breakthrough is, of course, the proposal \cite{ABJM,HLLLP2,ABJ} that the world-volume theory of $\min(N_1,N_2)$ M2-branes and $|N_2-N_1|$ fractional M2-branes on the orbifold ${\mathbb C}^4/{\mathbb Z}_k$ is described by the ${\mathcal N}=6$ supersymmetric Chern-Simons theory with the gauge group U$(N_1)_k\times$U$(N_2)_{-k}$ and two pairs of bifundamental matters.
Then, due to the localization technique \cite{P,KWY}, the partition function and the vacuum expectation values of the half-BPS Wilson loop operator on $S^3$, which are originally defined with the infinite-dimensional path integral, is reduced to a finite-dimensional matrix integration.
It is interesting to observe that the matrix model has a hidden structure of the gauge symmetry in the supergroup U$(N_1|N_2)$ \cite{GW,DT,MPtop}.

Another interesting progress is the study of this matrix model.
After the study of the large $N$ limit in the 't Hooft expansion \cite{DMP1,HKPT,DMP2}, where the degrees of freedom $N^{3/2}$ of the M2-branes were reproduced, it was found that all of the perturbative corrections in the large $N$ limit are summed up to the Airy function \cite{FHM,KEK}.
These studies further lead beautifully to an unexpected description of the Fermi gas formalism \cite{MP} where the partition function is reexpressed as that of a non-interacting Fermi gas system with a non-trivial one-particle Hamiltonian and the Chern-Simons level $k$ is identified as the Planck constant.
In the Fermi gas formalism, the behavior of the Airy function was reproduced in a few line computations, which immediately indicates the importance of the formalism.
The Fermi gas formalism was further used to study the non-perturbative effects in this matrix model with the WKB expansion \cite{MP,CM} and the exact values \cite{HMO1,PY,HMO2,HMO3}.
Finally, after combining these studies, it was proposed that the partition function \cite{HMMO} and the one-point function of the half-BPS Wilson loop operator \cite{HHMO} are respectively given by the free energy of the closed topological string theory and the open topological string theory on local ${\mathbb P}^1\times{\mathbb P}^1$.
The proposal was originally made for the case of equal ranks $N_2=N_1$ and later turned out to be valid in the rank deformation $N_2\ne N_1$ \cite{MM,HO} by generalizing the Fermi gas formalism.
(See \cite{PTEP,Marino} for reviews.)
One of the generalizations is called open string formalism \cite{HHMO,MM} and the other is called closed string formalism \cite{AHS,H,HO,Hosp,MS2,MN5,KM}.

There are several natural questions related to these developments.
First, so far we have only considered the partition function or the one-point function of the half-BPS Wilson loop operator, it is natural to ask whether we can generalize our analysis to more general correlations functions.
On one hand, in general, when two half-BPS Wilson loop operators preserve completely different supersymmetries, as a whole the correlation function does not preserve any supersymmetries at all, which prevents us from applying many techniques.
Especially since the localization technique for supersymmetric correlation functions does not work, our correlation function does not reduce to a matrix model any more. 
On the other hand, it is obvious that, for two identical half-BPS Wilson loops with only representations being possibly different, the two-point function reduces to a matrix model with the product of two characters.
Then, due to the Littlewood-Richardson rule, the products of two characters can be decomposed trivially into a linear combination of characters.
Hence, the correlation function with more than one insertion is not a new quantity but reduce to the one-point functions.
We hope to study two-point functions with a non-trivial and at the same time tractable structure.

Secondly, it was known that the partition function and the one-point functions enjoy many non-trivial relations, such as the Wronskian identity \cite{GHM2}, the open-closed duality \cite{HaOk,KM}, the Giambelli identity \cite{HHMO,MM,MaMo}, the Jacobi-Trudi identity \cite{FM} and so on.
It would be great if we can introduce a larger framework to combine all of the identities.

Thirdly, in the open string formalism, the one-point function of the half-BPS Wilson loop is given by a minor determinant of an infinite-dimensional matrix which contains two ingredients $H$ and $K$ (see \eqref{oneptFG}).
One of them $H$ is exactly the same as the one-point function of the half-BPS Wilson loop in the hook representation.
Although a combination of the other ingredients $K$ plays an important role especially in studying the partition function with the rank deformation \cite{MM,MNN}, the interpretation for a single component of it is missing.

Fourthly, in discussing the integrable structure of the ABJM matrix model \cite{MaMo,FM}, the starting point is the open string formalism \cite{MM}, where, unlike the indices of $H$, one of the indices of $K$ is always negative and appears consecutively.
It is natural to ask whether, in studying other correlation functions, we encounter a totally general minor determinant.

Fifthly, in studying the one-point function of the $1/6$-BPS Wilson loop operators \cite{KMSS,O6}, we encounter an imaginary contribution which cannot be regarded as a simple phase factor as that of the half-BPS Wilson loops.
Since we do not have much experience in the correlation functions containing imaginary parts, the study is difficult in general.
We hope to have more tractable examples of correlation functions with imaginary parts.

It turns out that all of these dissatisfactions can be alleviated by considering a certain type of two-point functions in the ABJM matrix model.
Namely, instead of introducing the characters both by $s_Y(e^\mu|e^\nu)$ and $s_Z(e^\mu|e^\nu)$, we invert the ``charge'' of one Wilson loop operator, and introduce a two-point function with the insertions $s_Y(e^\mu|e^\nu)$ and $s_Z(e^{-\mu}|e^{-\nu})$.

Of course it is natural to ask whether our capricious inversion of charges is physically relevant.
Especially we would like to know whether we can insert two different half-BPS Wilson loop operators in the ABJM theory preserving the total supersymmetries, and after applying the localization techniques, whether the insertion of these operators results in the two-point function in the matrix model we have introduced.
Although we do not have a concrete analysis to justify our expectation, we believe that this is possible due to the following arguments.
Since the scalar fields come into the half-BPS Wilson loop operator with the norm of the coordinates \cite{DT}, we believe that we can simultaneously reverse the sign of the scalar fields in the Wilson loop and the orientation of the loop to preserve the supersymmetries.
We hope that, after applying the localization techniques, the correlation function with these two insertions results in the two-point function we have defined.
Also, as we explain later, from the viewpoint of the Fermi gas formalism of the matrix model, we can still construct an open string formalism for the insertion of two characters with the opposite charges in a parallel manner as the one-point function in \cite{MM}.
The final result for the Fermi gas formalism of the two-point function is reminiscent of the representation of the supergroup U$(N_1|N_2)$ which is characterized by the so-called composite Young diagram \cite{Moens} combining two Young diagrams in the opposite directions.
Our result that the two-point function fully respects the hidden structure of the supergroup may suggest the naturalness of the definition and their origin in the ABJM theory.
Note also that this Fermi gas formalism of the two-point function generalizes that of the one-point function and therefore provides a larger framework.

After the introduction of the matrix model with two insertions of the opposite charges, we continue to study this two-point function.
Although it looks non-trivial at the beginning, after a long numerical analysis of the matrix model, we have found that the two-point function with the opposite charges is directly related to the two-point function with the same charges by the complex conjugate, which are subsequently related to the one-point functions with the Littlewood-Richardson rule.
We shall refer to this relation as a conjugate relation.

We also find an interesting relation for the imaginary part.
We find that, after the removal of the main phase factor, the imaginary part of the two-point function in the representations $Y$ and $Z$ reduces to a sum of the one-point functions in the representation $X$ whose box number is less than the sum of the box numbers of $Y$ and $Z$ by two.
Since we are studying the imaginary part of the two-point function and the result reduces to simpler quantities, this property may remind us of an interference between two insertions $Y$ and $Z$.
We shall refer to this relation as a descent relation.

Since the two-point function is related to the one-point function, on occasion of many numerical data, we also revisit the one-point function.
Although so far only the so-called diagonal BPS indices are identified which correspond to the case of equal ranks $N_2=N_1$, with the various numerical data in the rank deformation, we can investigate how the diagonal BPS indices are split by the degree difference.
Interestingly, we have found an asymmetry of the BPS indices in exchanging the two degrees $(d_+,d_-)$. 

This paper is organized as follows.
In the next section, we introduce the two-point function.
After establishing the Fermi gas formalism to study the two-point function, we proceed to studying it and find a few relations to the one-point functions.
In section \ref{onepoint}, we revisit the one-point function and investigate how the diagonal BPS indices are split.
Finally we conclude in section \ref{conclusion}.
In appendix \ref{determinant} we collect a few determinant formulas necessary for constructing the Fermi gas formalism of the two-point function and in appendix \ref{lowest} we compute the non-vanishing two-point function in the lowest rank.
In appendix \ref{superpose} we list a few data to study the conjugate relation, while appendix \ref{interfere} is devoted to the descent relation.
After presenting some relations and formulas for the one-point function in appendix \ref{onept} and appendix \ref{pert1pt}, in appendix \ref{np1pt} we list our exact expression of the non-perturbative part of the one-point function to study the split of the diagonal BPS indices.

\section{Two-point function}

In this section we shall introduce the two-point function in the ABJM matrix model, establish the Fermi gas formalism for it and study its property.

\subsection{Definition}\label{definition}

The one-point function of the half-BPS Wilson loop operator in the ABJM theory is reduced to a matrix model, after applying the localization technique for the supersymmetric theories \cite{P,KWY}\footnote{We follow the phase factor introduced in \cite{DMP1}.
This phase factor simplifies later formulas.},
\begin{align}
\langle s_Y\rangle_k(N_1,N_2)
&=i^{-\frac{1}{2}(N_1^2-N_2^2)}
\int\frac{D^{N_1}\mu}{N_1!}\frac{D^{N_2}\nu}{N_2!}
\frac{\prod_{m<m'}^{N_1}(2\sinh\frac{\mu_m-\mu_{m'}}{2})^2
\prod^{N_2}_{n<n'}(2\sinh\frac{\nu_n-\nu_{n'}}{2})^2}
{\prod_m^{N_1}\prod_n^{N_2}(2\cosh\frac{\mu_m-\nu_n}{2})^2}\nonumber\\
&\qquad\qquad\qquad\qquad\qquad\qquad\times
s_Y(e^\mu|e^\nu),
\end{align}
with
\begin{align}
D\mu=\frac{d\mu}{2\pi}e^{\frac{ik}{4\pi}\mu^2},\quad
D^{N_1}\mu=\prod_{m=1}^{N_1}D\mu_m,\quad
D\nu=\frac{d\nu}{2\pi}e^{-\frac{ik}{4\pi}\nu^2},\quad
D^{N_2}\nu=\prod_{n=1}^{N_2}D\nu_n.
\label{DmuDnu}
\end{align}
Here it was known that the hyperbolic functions can be regarded as a hyperbolic deformation of the invariant measure for the supergroup U$(N_1|N_2)$ and the Fresnel exponential factor can be regarded as the supertrace (see \cite{PTEP} for a review). 
Also the character $s_Y(e^\mu|e^\nu)$ is the super Schur polynomial, the character of the supergroup U$(N_1|N_2)$, and the arguments are the abbreviation, $s_Y(e^\mu|e^\nu)=s_Y(e^{\mu_1},e^{\mu_2},\cdots,e^{\mu_{N_1}}|e^{\nu_1},e^{\nu_2},\cdots,e^{\nu_{N_2}})$.

Although we do not have a rigorous localization analysis for two-point functions of the half-BPS Wilson loop operators in the ABJM theory, it is interesting to ask how we can define a two-point function naturally at the level of the matrix model.
If we simply insert another character $s_{Z}(e^\mu|e^\nu)$ in addition to the original one $s_{Y}(e^\mu|e^\nu)$ with the same arguments, the multiplication can be computed by the Littlewood-Richardson rule
\begin{align}
s_{Y}(e^\mu|e^\nu)s_{Z}(e^\mu|e^\nu)=\sum_{X}N^X_{YZ}s_{X}(e^\mu|e^\nu),
\end{align}
and the result reduces to the one-point function trivially.
Here let us consider the insertion of $s_{Z}(e^{-\mu}|e^{-\nu})=s_Z(e^{-\mu_1},e^{-\mu_2},\cdots,e^{-\mu_{N_1}}|e^{-\nu_1},e^{-\nu_2},\cdots,e^{-\nu_{N_2}})$ with the opposite ``charges''.

The insertion of the character with the opposite charges is partially motivated by the study of the Hopf links in the Chern-Simons matrix model \cite{AKMV}.\footnote{See also \cite{Kimura} for discussions on the similarity.}
The Chern-Simons theory is a topological theory, where the topological invariant of knots can be regarded as the correlation function of the Wilson loop operators \cite{WCS} and expressed as the Chern-Simons matrix model \cite{MCS}.
Especially for the Hopf links the matrix model is constructed by gluing two wave functions on the solid tori with the Wilson loop inside.
In the gluing process we effectively invert the charges of one of the Wilson loops.

After these discussions, let us define the two-point function in the ABJM matrix model as
\begin{align}
\langle s_Y\bar s_Z\rangle_k(N_1,N_2)
&=i^{-\frac{1}{2}(N_1^2-N_2^2)}
\int\frac{D^{N_1}\mu}{N_1!}\frac{D^{N_2}\nu}{N_2!}
\frac{\prod_{m<m'}^{N_1}(2\sinh\frac{\mu_m-\mu_{m'}}{2})^2
\prod^{N_2}_{n<n'}(2\sinh\frac{\nu_n-\nu_{n'}}{2})^2}
{\prod_m^{N_1}\prod_n^{N_2}(2\cosh\frac{\mu_m-\nu_n}{2})^2}\nonumber\\
&\qquad\qquad\qquad\qquad\qquad\qquad\times
s_Y(e^\mu|e^\nu)s_Z(e^{-\mu}|e^{-\nu}),
\label{two}
\end{align}
and the matrix model in the grand canonical ensemble as\footnote{It is important to match the power of $z$ with one of the ranks for both positive and negative $M$ \cite{MNN}.}
\begin{align}
\langle s_Y\bar s_Z\rangle^\text{GC}_{k,M}(z)
=\sum_{N=\max(0,-M)}^\infty z^N\langle s_Y\bar s_Z\rangle_k(N,N+M).
\label{gc}
\end{align}
Although our definition of the two-point function is not based on a concrete physical argument from the localization techniques, we shall see in the next subsection that this is in fact a nice definition which naturally incorporates the structure of the representation of the supergroup U$(N_1|N_2)$.

Before proceeding to the study of the two-point function, here let us shortly comment on their symmetries and relations.
First, we note that the two-point functions satisfy
\begin{align}
\langle s_Z\bar s_Y\rangle_k(N_1,N_2)&=\langle s_Y\bar s_Z\rangle_k(N_1,N_2),\nonumber\\
\langle s_{Y^\text{T}}\bar s_{Z^\text{T}}\rangle_k(N_2,N_1)
&=[\langle s_Y\bar s_Z\rangle_k(N_1,N_2)]^*,
\end{align}
which can be respectively proved by inverting the signs of the integration variables $\mu$ and $\nu$ and by inverting the integration variables $\mu$ and $\nu$ and using the transposition relation $s_{Y^\text{T}}(y|x)=s_Y(x|y)$.
In terms of the grand canonical ensemble, the two relations read
\begin{align}
\langle s_Z\bar s_Y\rangle^\text{GC}_{k,M}(z)
&=\langle s_Y\bar s_Z\rangle^\text{GC}_{k,M}(z),
\nonumber\\
\langle s_{Y^\text{T}}\bar s_{Z^\text{T}}\rangle^\text{GC}_{k,-M}(z)
&=z^M[\langle s_Y\bar s_Z\rangle^\text{GC}_{k,M}(z)]^*,
\end{align}
where the complex conjugate does not apply to $z$, $z^*=z$.
Secondly, when one of the Wilson loops is trivial, the two-point function reduces to the one-point function
\begin{align}
\langle s_\varnothing\bar s_Y\rangle_k(N_1,N_2)
&=\langle\bar s_Y\rangle_k(N_1,N_2)
=\langle s_Y\rangle_k(N_1,N_2)
=\langle s_Y\bar s_\varnothing\rangle_k(N_1,N_2).
\end{align} 

\subsection{Fermi gas formalism}

In this subsection we construct the Fermi gas formalism to study this matrix model.
Although so far it was only noted that the case of $M=N_2-N_1<0$ can be studied by taking the complex conjugate, here we study the cases of $M\ge 0$ and $M\le 0$ separately and point out that they are connected smoothly at $M=0$.
In the both cases, the resulting Fermi gas formalism is schematically summarized by the expression
\begin{align}
\langle s_Y\bar s_{Y'}\rangle^\text{GC}_{k,M}(z)
=i^{\frac{1}{2}M^2}\Xi_k(w)\det\begin{pmatrix}
{\cal H}_k^{(\widetilde a|\widetilde l)}(w)
\end{pmatrix}_{(\widetilde a,\widetilde l)\in\widetilde A\times\widetilde L},
\label{FG}
\end{align}
where $w$ is related to $z$ by $w=(-i)^Mz$.

\begin{figure}[!ht]
\centering\includegraphics[scale=0.6,angle=-90]{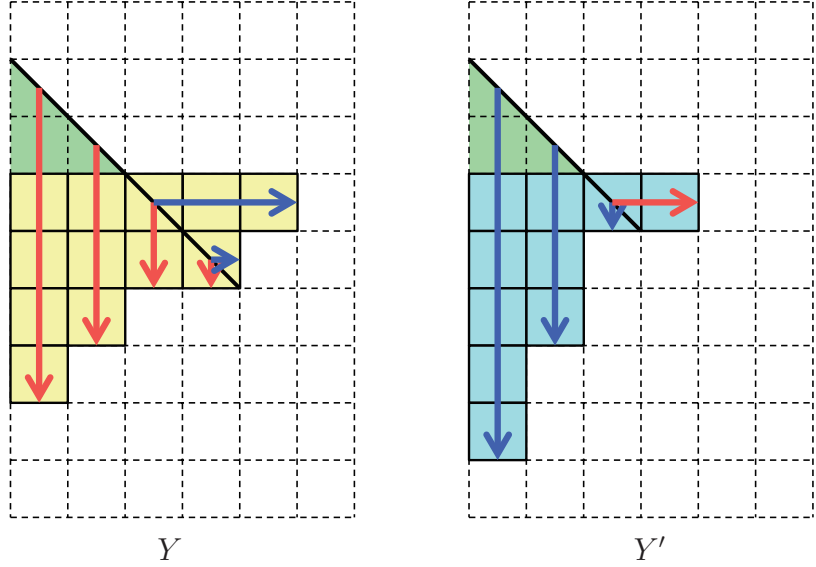}\\[6pt]
$Y$\hspace{60mm}$Y'$
\caption{The shifted Frobenius notation for $Y=[5,4,2,1]$ (left) and $Y'=[4,2,2,1,1]$ (right).
For $Y$ we adopt the original shifted Frobenius notation $(a_1,a_2|l_1,l_2,l_3,l_4)_{M=2}=(\frac{5}{2},\frac{1}{2}|\frac{11}{2},\frac{7}{2},\frac{3}{2},\frac{1}{2})_{M=2}$ defined in \eqref{al}.
For $Y'$ we invert the signs and the roles of arms and legs $(-l'_1|{-a'_1},-a'_2,-a'_3)_{M=2}=(\frac{3}{2}|\frac{13}{2},\frac{7}{2},\frac{1}{2})_{M=2}$ as in \eqref{la}.
}
\label{young}
\end{figure}

\begin{figure}[!ht]
\centering\includegraphics[scale=0.6,angle=-90]{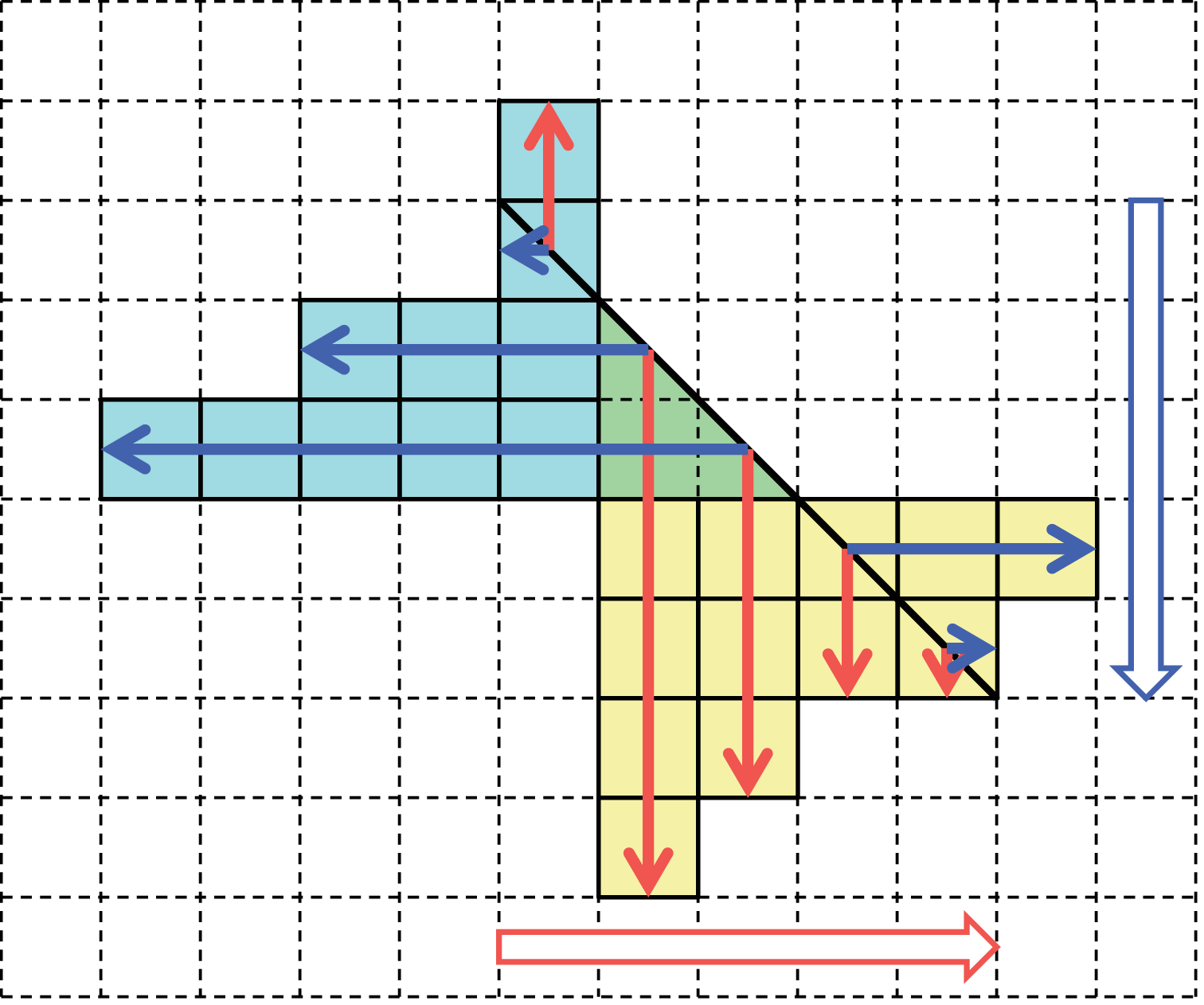}
\caption{After reversing the order of arms and legs in $Y'$, the arm and leg lengths in the Fermi gas formalism \eqref{FG} are given by $(a'_3,a'_2,a'_1,a_1,a_2|l'_1,l_1,l_2,l_3,l_4)=(-\frac{1}{2},-\frac{7}{2},-\frac{13}{2},\frac{5}{2},\frac{1}{2}|{-\frac{3}{2}},\frac{11}{2},\frac{7}{2},\frac{3}{2},\frac{1}{2})$ as in \eqref{MshiftedYZ}.}
\label{composite}
\end{figure}

Before explaining various quantities, let us first explain the notation of the Young diagram and the structure of the determinant in \eqref{FG}.
See figure \ref{young} for examples.
Here we denote the Young diagrams $Y$ by the so-called $M$-shifted Frobenius notation
\begin{align}
Y=(a_1,a_2,\cdots,a_R|l_1,l_2,\cdots,l_{M+R})_M,\quad
R=\max\{i|a_i>0\}=\max\{j|l_j>0\}-M.
\label{MshiftedY}
\end{align}
This is obtained by listing the horizontal lengths and the vertical lengths from the diagonal line shifted by $M$ upward/rightward (or $|M|$ downward/leftward when $M<0$) to the boundary of the Young diagram as arm lengths and leg lengths, which are also expressed as
\begin{align}
a_i=\lambda_i-i+\frac{1}{2}-M,\quad l_j=\lambda^\text{T}_j-j+\frac{1}{2}+M,
\label{al}
\end{align}
with the standard notation of the Young diagram by listing the non-vanishing numbers of horizontal boxes $Y=[\lambda_1,\lambda_2,\cdots]$ or dually $Y^\text{T}=[\lambda^\text{T}_1,\lambda^\text{T}_2,\cdots]$.
The Young diagram $Y'$ is also denoted by the $M$-shifted Frobenius notation, though we invert the signs and the roles of the arm lengths and the leg lengths
\begin{align}
Y'=(-l'_1,\cdots,-l'_{R'}|{-a'_1},\cdots,-a'_{M+R'})_M,\quad
R'=\max\{j|l'_j<0\}=\max\{i|a'_i<0\}-M.
\label{Y'al}
\end{align}
with
\begin{align}
-l'_j=\lambda'_j-j+\frac{1}{2}-M,\quad
-a'_i=\lambda'^{\text{T}}_i-i+\frac{1}{2}+M,
\label{la}
\end{align}
for $Y'=[\lambda'_1,\lambda'_2,\cdots]$.
The column indices and the row indices appearing in the determinant in \eqref{FG} are
\begin{align}
\widetilde A=\{a'_{R'+M},a'_{R'+M-1},\cdots,a'_1,a_1,a_2,\cdots,a_R\},\quad
\widetilde L=\{l'_{R'},l'_{R'-1},\cdots,l'_1,l_1,l_2,\cdots,l_{M+R}\},
\label{MshiftedYZ}
\end{align}
respectively.
Note that, compared with \eqref{Y'al}, in \eqref{MshiftedYZ} the signs and the roles of the arm lengths and the leg lengths in the Young diagram $Y'$ are exchanged to combine with those in $Y$.
This is why we have introduced the notation in \eqref{Y'al}.
Pictorially, this exchange is clearly encoded by reversing one of the Young diagrams $Y'$.
See figure \ref{composite} for an explanation for the example of the two Young diagrams $Y$ and $Y'$ given in figure \ref{young}.
The diagram combining two Young diagrams in the opposite directions is called the composite Young diagram and appear naturally in the study of the representation of the supergroup U$(N_1|N_2)$ \cite{Moens}.
We shall refer to the order of the arm lengths and the leg lengths in \eqref{MshiftedY} as the standard order for a single Young diagram and the order in \eqref{MshiftedYZ} as the standard order for a composite Young diagram.

Although at the first sight the matrix $\bigl({\cal H}_k^{(\widetilde a|\widetilde l)}(w)\bigr)_{\widetilde A\times\widetilde L}$ in \eqref{FG} consists of four blocks respectively with positive/negative arm/leg lengths, they reduce essentially to two blocks. 
If we refer to the matrix elements as $H_k^{(a|l)}$ and $K_k^{(a'|l)}$ when the leg length is positive, the matrix elements for the negative leg length are also given in terms of $H_k^{(a|l)}$ and $K_k^{(a'|l)}$ by the complex conjugate as
\begin{align}
\Bigl({\cal H}_k^{(\widetilde a|\widetilde l)}(w)\Bigr)_{\widetilde A\times\widetilde L}
=\begin{pmatrix}
\bigl[H_k^{(-a'|-l')}(w)\bigr]^*_{(R'+M)\times R'}&
\bigl[K_k^{(a'|l)}(w)\bigr]_{(R'+M)\times(M+R)}\\
\bigl[-wK_k^{(-a|-l')}(w)\bigr]^*_{R\times R'}&
\bigl[H_k^{(a|l)}(w)\bigr]_{R\times(M+R)}
\end{pmatrix},
\label{calH}
\end{align}
where the complex conjugate does not apply to $w$, $w^*=w$.
Of course, as in the left-hand side, the indices on the right-hand side are also given in the standard order of the composite Young diagram \eqref{MshiftedYZ}. 
Various quantities are schematically defined as
\begin{align}
\Xi_k(w)&=\Det(1+wPQ),\nonumber\\
K_k^{(a'|l)}(w)&=F_l(1+wQP)^{-1}F_{a'},\nonumber\\
H_k^{(a|l)}(w)&=wF_l(1+wQP)^{-1}QE_a,
\label{ingredients}
\end{align}
with
\begin{align}
P(\mu,\nu)=\frac{1}{2\cosh\frac{\mu-\nu}{2}},\quad
Q(\nu,\mu)=\frac{1}{2\cosh\frac{\nu-\mu}{2}},\quad
E_p(\mu)=e^{p\mu},\quad
F_q(\nu)=e^{q\nu}.
\end{align}
Here we regard $P(\mu,\nu)$, $Q(\nu,\mu)$ as matrices and $E_p(\mu)$, $F_q(\nu)$ as vectors.
When multiplying by contracting the ``indices'' $\mu$ and $\nu$ we utilize the integration measure \eqref{DmuDnu}.
The determinant $\Det$ used in defining $\Xi_k(w)$ in \eqref{ingredients} is the Fredholm determinant on the function space.

Note that, for the partition function or the one-point function, the expression \eqref{FG} reduces to those found in \cite{MP,HHMO,MM}.
Especially, for the grand canonical partition function of equal ranks $N_2=N_1$ or $M=0$ \cite{MP}, we have
$\langle 1\rangle^\text{GC}_{k,M=0}(z)
=\Xi_k(w)$, which gives a physical interpretation to $\Xi_k(w)$.
For the one-point function of the Wilson loop operator in the hook representation for the case of equal ranks $N_2=N_1$ \cite{HHMO}, we find
$\langle s_{(a|l)}\rangle^\text{GC}_{k,M=0}(z)
=\Xi_k(w)H_k^{(a|l)}(w)$.
Hence, after the normalization, $H_k^{(a|l)}(w)$ is clearly interpreted as the one-point function in the hook representation.
For the grand canonical partition function with the rank deformation $N_2>N_1$ or $M>0$ \cite{MM}, we find
$\langle 1\rangle^\text{GC}_{k,M}(z)
=i^{\frac{1}{2}M^2}\Xi_k(w)\det[K_k^{(a'|l)}(w)]_{M\times M}$.
Though the determinant is important in computing the grand canonical partition function with the rank deformation, the interpretation of the single component of $K_k^{(a'|l)}(w)$ is not clear.
After our introduction of the two-point function, the situation is improved.
We can consider, for example, the two-point function
\begin{align}
\langle s_{(\frac{1}{2}|l)}\bar s_{(\frac{1}{2}|-a')}\rangle^\text{GC}_{k,M=1}(z)
=i^{\frac{1}{2}}\Xi_k(w)K_k^{(a'-1|l+1)}(w),
\end{align}
which gives the interpretation for the single component of $K_k^{(a'|l)}(w)$.

Also note that, for the general one-point function of the Wilson loop operator with the rank deformation $M>0$, we find \cite{MM}
\begin{align}
\langle s_Y\rangle^\text{GC}_{k,M}(z)
=i^{\frac{1}{2}M^2}\Xi_k(w)\det\begin{pmatrix}
\bigl[K_k^{(a'_i|l_j)}(w)\bigr]_{M\times(M+R)}\\
\bigl[H_k^{(a_i|l_j)}(w)\bigr]_{R\times(M+R)}
\end{pmatrix}.
\label{oneptFG}
\end{align}
Regarding the determinant as a minor determinant of an infinite-dimensional matrix, the matrix element with negative leg lengths never appears.
Besides, although the indices of $a_i$ and $l_j$ depend on the shape of the Young diagram and can appear generally by choosing different Young diagrams, the indices of $a'_i$ always range consecutively within
\begin{align}
a'_i\in\Bigl\{-\frac{1}{2},-\frac{3}{2},\cdots,-M+\frac{1}{2}\Bigr\}.
\label{contarms}
\end{align}
The situation is not improved much even we include the case of $M<0$.
This may imply that the one-point function is not the most general quantity to study and there is a natural generalization of it.
After obtaining the Fermi gas formalism for the general two-point function, the matrix elements with both of the lengths in $(\widetilde a|\widetilde l)$ being negative can participate and the indices $a'$ and $l'$ do not have to range consecutively any more.

\subsection{Derivation}\label{derivation}

In this subsection, we shall give a derivation of the Fermi gas formalism for the two-point function \eqref{FG} from the definition \eqref{two}.
The basic techniques for the derivation already appeared in the derivation for the one-point function \cite{MM}.
We first introduce two determinant formulas, where one is used to express the integration measure in a determinant and the other is used to express the super Schur polynomial by replacing the previous determinant by another.
Then we can combine these two determinants by the continuous Cauchy-Binet formula.
The only new ingredient for the two-point function is to repeat these techniques twice.

\subsubsection{$M\ge 0$}

We first explain the Fermi gas formalism for the case of $N_2\ge N_1$ carefully.
Namely, we set $N_1=N$ and $N_2=N+M$ with $M\ge 0$.
For the integration measure, we introduce the combination of the Vandermonde determinant and the Cauchy determinant
\begin{align}
\frac{\prod_{m<m'}^{N_1}(x_m-x_{m'})
\prod_{n<n'}^{N_2}(y_n-y_{n'})}
{\prod_{m=1}^{N_1}\prod_{n=1}^{N_2}(x_m+y_n)}
=(-1)^{N_1(N_2-N_1)}
\det\begin{pmatrix}\biggl[\displaystyle\frac{1}{x_m+y_n}\biggr]
_{(m,n)\in Z_{1}\times Z_{2}}\\
\Bigl[y_n^{\overline l-\frac{1}{2}}\Bigr]
_{(\overline l,n)\in\overline L\times Z_{2}}
\end{pmatrix},
\label{Z}
\end{align}
with $Z_1=\{1,2,\cdots,N_1\}$, $Z_2=\{1,2,\cdots,N_2\}$ and 
$\overline L=\{M-\frac{1}{2},M-\frac{3}{2},\cdots,\frac{1}{2}\}$ appearing in the determinant in this order.
For the super Schur polynomial, we utilize the determinant formula \cite{MVdJ}
\begin{align}
s_Y(x|y)
&=(-1)^R\det\begin{pmatrix}\biggl[\displaystyle\frac{1}{x_m+y_n}\biggr]
_{(m,n)\in Z_{1}\times Z_{2}}&
\Bigl[x_m^{a-\frac{1}{2}}\Bigr]
_{(m,a)\in Z_{1}\times A}\\
\Bigl[y_n^{l-\frac{1}{2}}\Bigr]
_{(l,n)\in L\times Z_{2}}&
[0]_{L\times A}\end{pmatrix}\nonumber\\
&\qquad\qquad\bigg/
\det\begin{pmatrix}\biggl[\displaystyle\frac{1}{x_m+y_n}\biggr]
_{(m,n)\in Z_{1}\times Z_{2}}\\
\Bigl[y_n^{\overline l-\frac{1}{2}}\Bigr]
_{(\overline l,n)\in\overline L\times Z_{2}}\end{pmatrix},
\label{Y}
\end{align}
where $A=\{a_1,a_2,\cdots,a_R\}$ and $L=\{l_1,l_2,\cdots,l_{M+R}\}$ are the sets of the arm and leg lengths in the shifted Frobenius notation in the standard order \eqref{MshiftedY}.

Since we have the square in the integration measure and two super Schur functions in the two-point function \eqref{two}, we need a copy of the previous two determinants.
It is easier for the later convenience to obtain them by substituting $x\to x^{-1}$, $y\to y^{-1}$, replacing $Y$ by $Y'$ (namely $\overline l\to -\overline a$, $a\to -l'$, $l\to -a'$ as in \eqref{Y'al}) and transposing the matrix.
Then, we find
\begin{align}
&(-1)^{\frac{1}{2}N_1(N_1-1)+\frac{1}{2}N_2(N_2-1)}
\frac{\prod_{m<m'}^{N_1}(x_{m'}^{-1}-x_m^{-1})
\prod_{n<n'}^{N_2}(y_{n'}^{-1}-y_n^{-1})}
{\prod_{m=1}^{N_1}\prod_{n=1}^{N_2}(x_m^{-1}+y_n^{-1})}\nonumber\\
&=(-1)^{N_1(N_2-N_1)}
\det\begin{pmatrix}\biggl[\displaystyle\frac{1}{y_n^{-1}+x_m^{-1}}\biggr]
_{(n,m)\in Z_{2}\times Z_{1}}&
\Bigl[y_n^{\overline a+\frac{1}{2}}\Bigr]
_{(n,\overline a)\in Z_{2}\times\overline A}
\end{pmatrix},
\label{Zbar}
\end{align}
with $\overline A=\{-(M-\frac{1}{2}),-(M-\frac{3}{2}),\cdots,-\frac{1}{2}\}$ and 
\begin{align}
s_{Y'}(x^{-1}|y^{-1})
&=(-1)^{R'}\det\begin{pmatrix}\biggl[\displaystyle\frac{1}{y_n^{-1}+x_m^{-1}}\biggr]
_{(n,m)\in Z_{2}\times Z_{1}}&
\Bigl[y_n^{a'+\frac{1}{2}}\Bigr]
_{(n,a')\in Z_{2}\times A'}\\
\Bigl[x_m^{l'+\frac{1}{2}}\Bigr]
_{(l',m)\in L'\times Z_{1}}&
[0]_{L'\times A'}\end{pmatrix}\nonumber\\
&\qquad\bigg/
\det\begin{pmatrix}\biggl[\displaystyle\frac{1}{y_n^{-1}+x_m^{-1}}\biggr]
_{(n,m)\in Z_{2}\times Z_{1}}&
\Bigl[y_n^{\overline a+\frac{1}{2}}\Bigr]
_{(n,\overline a)\in Z_{2}\times\overline A}\end{pmatrix},
\label{Ybar}
\end{align}
where $-L'=\{-l'_1,-l'_2,\cdots,-l'_{R'}\}$ and $-A'=\{-a'_1,-a'_2,\cdots,-a'_{M+R'}\}$ are respectively the sets of the {\it arm} and {\it leg} lengths for the single Young diagram $Y'$.
See figure \ref{young} again to avoid confusion.

After substituting the four determinant formulas \eqref{Z}, \eqref{Y}, \eqref{Zbar}, \eqref{Ybar} with $x_m=e^{\mu_m}$, $y_n=e^{\nu_n}$ and multiplying them all together, finally we find that the two-point function is given by
\begin{align}
&\langle s_Y\bar s_{Y'}\rangle_{k}(N,N+M)
=i^{-\frac{1}{2}(N_1^2-N_2^2)}(-1)^{\frac{1}{2}N_1(N_1-1)+\frac{1}{2}N_2(N_2-1)+R+R'}
\int\frac{D^{N_1}\mu}{N_1!}\frac{D^{N_2}\nu}{N_2!}\nonumber\\
&\times\det\begin{pmatrix}
[P(\mu,\nu)]_{N\times(N+M)}
&[E_a(\mu)]_{N\times R}\\
[F_l(\nu)]_{(M+R)\times(N+M)}
&[0]_{(M+R)\times R}
\end{pmatrix}
\det\begin{pmatrix}
[Q(\nu,\mu)]_{(N+M)\times N}
&[F_{a'}(\nu)]_{(N+M)\times(M+R')}\\
[E_{l'}(\mu)]_{R'\times N}
&[0]_{R'\times(M+R')}
\end{pmatrix},
\label{detdet}
\end{align}
with
\begin{align}
&P(\mu,\nu)=\frac{1}{2\cosh\frac{\mu-\nu}{2}},\quad
E_a(\mu)=e^{a\mu},\quad
F_l(\nu)=e^{l\nu},\nonumber\\
&Q(\nu,\mu)=\frac{1}{2\cosh\frac{\nu-\mu}{2}},\quad
F_{a'}(\nu)=e^{a'\nu},\quad
E_{l'}(\mu)=e^{l'\mu}.
\end{align}

Now using the continuous Cauchy-Binet formula, {\it Formula 1} given in appendix \ref{determinant}, we can combine two determinants into
\begin{align}
&\langle s_Y\bar s_{Y'}\rangle_{k,M}(N)
=i^{MN+\frac{1}{2}M^2}(-1)^{MN+\frac{1}{2}M(M-1)+R+R'+RR'}
\int\frac{D^N\mu}{N!}\nonumber\\
&\quad\times\det\begin{pmatrix}
[(P\circ Q)(\mu,\mu)]_{N\times N}
&[(P\circ F_{a'})(\mu)]_{N\times(M+R')}
&[E_a(\mu)]_{N\times R}\\
[(F_l\circ Q)(\mu)]_{(M+R)\times N}
&[(F_l\circ F_{a'})]_{(M+R)\times(M+R')}
&[0]_{(M+R)\times R}\\
[E_{l'}(\mu)]_{R'\times N}
&[0]_{R'\times(M+R')}
&[0]_{R'\times R}
\end{pmatrix},
\end{align}
with $\circ$ denoting the contraction with $D\nu$ \eqref{DmuDnu}.
Using {\it Formula 2}, we can express the two-point function in the grand canonical ensemble defined in \eqref{gc} as
\begin{align}
\langle s_Y\bar s_{Y'}\rangle^\text{GC}_{k,M}(z)
=i^{\frac{1}{2}M^2}(-1)^{\frac{1}{2}M(M-1)+R+R'+RR'}
\Det\begin{pmatrix}1+wP\circ Q&wP\circ F_{a'}&wE_a\\
F_l\circ Q&F_l\circ F_{a'}&0\\E_{l'}&0&0\end{pmatrix}.
\end{align}
Here we have introduced $w=(-i)^Mz$ to take care of the sign factor proportional to $N$.
The determinant $\Det$ is a combination of the Fredholm determinant for the first block of rows and columns and the usual determinant for the remaining blocks.
Using {\it Formula 3} we can further rewrite it as
\begin{align}
&\langle s_Y\bar s_{Y'}\rangle^\text{GC}_{k,M}(z)
=i^{\frac{1}{2}M^2}(-1)^{\frac{1}{2}M(M-1)+R+R'+RR'}
\Det(1+wQP)\nonumber\\
&\quad\times
\det\begin{pmatrix}[F_l(1+wQP)^{-1}F_{a'}]_{(M+R)\times(M+R')}&
[-wF_l(1+wQP)^{-1}QE_a]_{(M+R)\times R}\\
[-wE_{l'}(1+wPQ)^{-1}PF_{a'}]_{R'\times(M+R')}&
[-wE_{l'}(1+wPQ)^{-1}E_a]_{R'\times R}
\end{pmatrix},
\end{align}
where we have dropped $\circ$ and understand the matrix multiplications by the integrations $D\mu$ and $D\nu$ tacitly.
Now it is very interesting to observe that the arm and leg lengths are those appearing in the composite Young diagram in figure \ref{composite}.

To reduce the expression, we first multiply the sign factors $(-1)^{R+R'}$ to the second column block and the second row block.
After that we exchange the first and second row block and rearrange the arm and leg lengths to the standard order of the composite Young diagram.
Due to the exchange of the rows and columns, we encounter the extra sign factors
\begin{align}
(-1)^{(M+R)R'+\frac{1}{2}R'(R'-1)+\frac{1}{2}(M+R')(M+R'-1)}=(-1)^{\frac{1}{2}M(M-1)+RR'},
\end{align}
cancelling parts of the sign factor.
Finally, the two-point function in the grand canonical ensemble is given by
\begin{align}
&\langle s_Y\bar s_{Y'}\rangle^\text{GC}_{k,M}(z)
=i^{\frac{1}{2}M^2}\Det(1+wQP)\nonumber\\
&\quad\times\det\begin{pmatrix}[wE_{l'}(1+w PQ)^{-1}PF_{a'}]_{R'\times(R'+M)}&
[-wE_{l'}(1+wPQ)^{-1}E_a]_{R'\times R}\\
[F_l(1+wQP)^{-1}F_{a'}]_{(M+R)\times(R'+M)}&
[wF_l(1+wQP)^{-1}QE_a]_{(M+R)\times R}\end{pmatrix}.
\label{fermi}
\end{align}
After transposing the determinant in this expression and using
\begin{align}
&wE_{l'}(1+wPQ)^{-1}PF_{a'}=wE_{-l'}(1+wPQ)^{-1}PF_{-a'}=[wF_{-l'}(1+wQP)^{-1}QE_{-a'}]^*,
\nonumber\\
&{-wE_{l'}(1+wPQ)^{-1}E_{a}}=-wE_{-l'}(1+wPQ)^{-1}E_{-a}=[-wF_{-l'}(1+wQP)^{-1}F_{-a}]^*,
\end{align}
we obtain \eqref{FG}.
Note that the complex conjugate can be realized effectively by exchanging the matrices $(P,Q)$ and the vectors $(E,F)$ simultaneously. 

\subsubsection{$M\le 0$}

The construction of the Fermi gas formalism for the case of $M\le 0$ does not change much.
Instead of keeping the same notation with $M\le 0$, let us introduce the notation $M=-\bar M$, $N_1=N=\bar N+\bar M$, $N_2=N+M=\bar N$ which is more intuitive.
This time instead of \eqref{Z} and \eqref{Zbar} we use
\begin{align}
&\frac{\prod_{m<m'}^{N_1}(x_m-x_{m'})
\prod_{n<n'}^{N_2}(y_n-y_{n'})}
{\prod_{m=1}^{N_1}\prod_{n=1}^{N_2}(x_m+y_n)}\nonumber\\
&\quad=(-1)^{N_2(N_1-N_2)}
\det\begin{pmatrix}\biggl[\displaystyle\frac{1}{x_m+y_n}\biggr]
_{(m,n)\in Z_{1}\times Z_{2}}&
\Bigl[x_m^{\overline a-\frac{1}{2}}\Bigr]
_{(m,\overline a)\in Z_{1}\times\overline A}
\end{pmatrix},
\label{ZMneg}
\end{align}
with $\overline A=\{\bar M-\frac{1}{2},\bar M-\frac{3}{2},\cdots,\frac{1}{2}\}$ and
\begin{align}
&(-1)^{\frac{1}{2}N_1(N_1-1)+\frac{1}{2}N_2(N_2-1)}
\frac{\prod_{m<m'}^{N_1}(x_{m'}^{-1}-x_m^{-1})
\prod_{n<n'}^{N_2}(y_{n'}^{-1}-y_n^{-1})}
{\prod_{m=1}^{N_1}\prod_{n=1}^{N_2}(x_m^{-1}+y_n^{-1})}\nonumber\\
&=(-1)^{N_2(N_1-N_2)}
\det\begin{pmatrix}\biggl[\displaystyle\frac{1}{y_n^{-1}+x_m^{-1}}\biggr]
_{(n,m)\in Z_{2}\times Z_{1}}\\
\Bigl[x_m^{\overline l+\frac{1}{2}}\Bigr]
_{(\overline l,m)\in\overline L\times Z_{1}}
\end{pmatrix},
\label{ZbarMneg}
\end{align}
with $\overline L=\{-(\bar M-\frac{1}{2}),-(\bar M-\frac{3}{2}),\cdots,-\frac{1}{2}\}$ and change the denominators of \eqref{Y} and \eqref{Ybar} accordingly.
Also for the $M$-shifted Frobenius notation of the Young diagram, we introduce
\begin{align}
Y&=(a_1,\cdots,a_{\bar M+\bar R}|l_1,\cdots,l_{\bar R})_{-\bar M},\quad
\bar R=\max\{i|a_i>0\}-\bar M=\max\{j|l_j>0\},\nonumber\\
Y'&=(-l'_1,\cdots,-l'_{\bar M+\bar R'}|-a'_1,\cdots,-a'_{\bar R'})_{-\bar M},\quad
\bar R'=\max\{j|l'_j<0\}-\bar M=\max\{i|a'_i<0\},
\end{align}
where we note that $R$ and $\bar R$ are related by $R=\bar M+\bar R$, $M+R=\bar R$ and $R'$ and $\bar R'$ are related by $R'=\bar M+\bar R'$, $M+R'=\bar R'$.
Then we arrive at the same expression as in \eqref{detdet}.
\begin{align}
&\langle s_Y\bar s_{Y'}\rangle_{k}(\bar N+\bar M,\bar N)
=i^{-\frac{1}{2}(N_1^2-N_2^2)}(-1)^{\frac{1}{2}N_1(N_1-1)+\frac{1}{2}N_2(N_2-1)+\bar R+\bar R'}
\int\frac{D^{N_1}\mu}{N_1!}\frac{D^{N_2}\nu}{N_2!}\nonumber\\
&\times\det\begin{pmatrix}
[P(\mu,\nu)]_{(\bar N+\bar M)\times\bar N}
&[E_a(\mu)]_{(\bar N+\bar M)\times(\bar M+\bar R)}\\
[F_l(\nu)]_{\bar R\times\bar N}
&[0]_{\bar R\times(\bar M+\bar R)}
\end{pmatrix}
\det\begin{pmatrix}
[Q(\nu,\mu)]_{\bar N\times(\bar N+\bar M)}
&[F_{a'}(\nu)]_{\bar N\times\bar R'}\\
[E_{l'}(\mu)]_{(\bar M+\bar R')\times(\bar N+\bar M)}
&[0]_{(\bar M+\bar R')\times\bar R'}
\end{pmatrix}.
\end{align}

This time, since we have more $\mu$ variables than $\nu$ variables, we shall perform the integration $D\mu$ first and then move to the grand canonical ensemble \eqref{gc}
\begin{align}
\langle s_Y\bar s_{Y'}\rangle^\text{GC}_{k,-\bar M}(z)
=\sum_{\bar N=0}^\infty z^{\bar N+\bar M}\langle s_Y\bar s_{Y'}\rangle_{k}(\bar N+\bar M,\bar N).
\end{align}
Effectively we can exchange the two determinants in the integrand and proceed in the parallel manner.
Finally, we find
\begin{align}
&\langle s_Y\bar s_{Y'}\rangle^\text{GC}_{k,-\bar M}(z)
=i^{\frac{1}{2}\bar M^2}
(-1)^{\frac{1}{2}\bar M(\bar M-1)+\bar R+\bar R'+\bar R\bar R'}
\Det(1+w QP)\nonumber\\
&\quad\times
(-w)^{\bar M}\det\begin{pmatrix}
[E_{l'}(1+wPQ)^{-1}E_a]_{(\bar M+\bar R')\times(\bar M+\bar R)}&
[-wE_{l'}(1+wPQ)^{-1}PF_{a'}]_{(\bar M+\bar R')\times\bar R'}\\
[-wF_l(1+wQP)^{-1}QE_a]_{\bar R\times(\bar M+\bar R)}&
[-wF_l(1+wQP)^{-1}F_{a'}]_{\bar R\times\bar R'}
\end{pmatrix}.
\end{align}
After changing the rows and the columns suitably and transposing the determinant, we arrive at the same expression as \eqref{fermi}.

\subsection{Phase factor}\label{secphase}

After establishing the Fermi gas formalism for the two-point function in the ABJM matrix model in the previous subsection, we can start the computation.
We first note that we can compute the lowest component of the grand canonical two-point function in the expansion of $z$ (in other words, the non-vanishing canonical two-point function $\langle s_Y\bar s_Z\rangle_k(N,N+M)$ of the lowest rank $N$) and present the results in terms of the Young diagram $Y$ and $Y'$.
The result is given in appendix \ref{lowest}.

For higher orders we need to perform the residue integration order by order for each ingredient \eqref{ingredients} appearing in \eqref{FG} and \eqref{calH}.
Fortunately, the computation of each part already appeared previously.
For $\Xi_k(w)$, the computation was already given in the first computation of the exact values for the partition function in \cite{HMO1,PY,HMO2}.
The techniques of rewriting the multiplications among matrices into subsequent multiplications by matrices on a vector were also reviewed in \cite{PTEP}.
For $H_k^{(a|l)}(w)$, the computation was given in the study of the one-point function of the Wilson loop in \cite{HHMO}, where it was found that the computation is convergent for $2(a+l)<k$.
For $K_k^{(a'|l)}(w)$, the computation was given in the study of the partition function with the rank deformation $N_2\ne N_1$ in \cite{MM}, where the convergence is valid for $2|a'|<k$ and $2l<k$.
Using these results of the computation, we can compute the two-point function without difficulty.
For simplicity in discussing the numerical results, we always consider the case of $M\ge 0$.
We have computed the two-point function $\langle s_Y\bar s_Z\rangle_{k}(N,N+M)$ of $2\le|Y|+|Z|\le 5$ and $k=3,4,6,8,12$ up to $N=N_\text{max}$ with $(k,N_\text{max})=(3,7),(4,13),(6,8),(8,4),(12,5)$ for $M$ within the range of convergence.

As a preliminary study of the result, we start with the phase factor.
As known in \cite{HHMO,MM}, the phase dependence of the partition function and the one-point function is rather trivial.
Especially, with the phase factor $i^{-\frac{1}{2}(N_1^2-N_2^2)}$ included in the definition of \eqref{two} \cite{DMP1}, the phase of the partition function $e^{i\Theta^\varnothing_{k,M}}$ and that of the one-point function $e^{i\Theta_{k,M}^Y}$ defined by
\begin{align}
\frac{\langle 1\rangle_{k}(N,N+M)}{|\langle 1\rangle_{k}(N,N+M)|}
=e^{i\Theta^\varnothing_{k,M}},\quad
\frac{\langle s_Y\rangle_{k}(N,N+M)}{|\langle s_Y\rangle_{k}(N,N+M)|}
=e^{i\Theta_{k,M}^Y},
\end{align}
are given by
\begin{align}
\Theta_{k,M}^Y=\theta_{k,M}+\theta_{k,M}^Y,\quad
\theta_{k,M}=-\frac{\pi}{k}\frac{M^3-M}{6},\quad
\theta_{k,M}^Y=\frac{\pi}{k}(2c^Y-M|Y|),
\label{1ptphase}
\end{align}
independent of the rank $N$.
Here $|Y|$ is the total box number of the Young diagram $Y$ and $c^Y$ is the sum of the contents for the Young diagram $Y$, 
\begin{align}
|Y|=\sum_{(i,j)\in Y}1,\quad
c^Y=\sum_{(i,j)\in Y}(j-i).
\label{contents}
\end{align}
Here we stress that, with the phase $i^{-\frac{1}{2}(N_1^2-N_2^2)}$ included, all of the remaining phases in \eqref{1ptphase} are proportional to $k^{-1}$ and cannot be removed simply by changing rows or columns.

\begin{figure}[!ht]
\begin{center}
\includegraphics[scale=0.6]{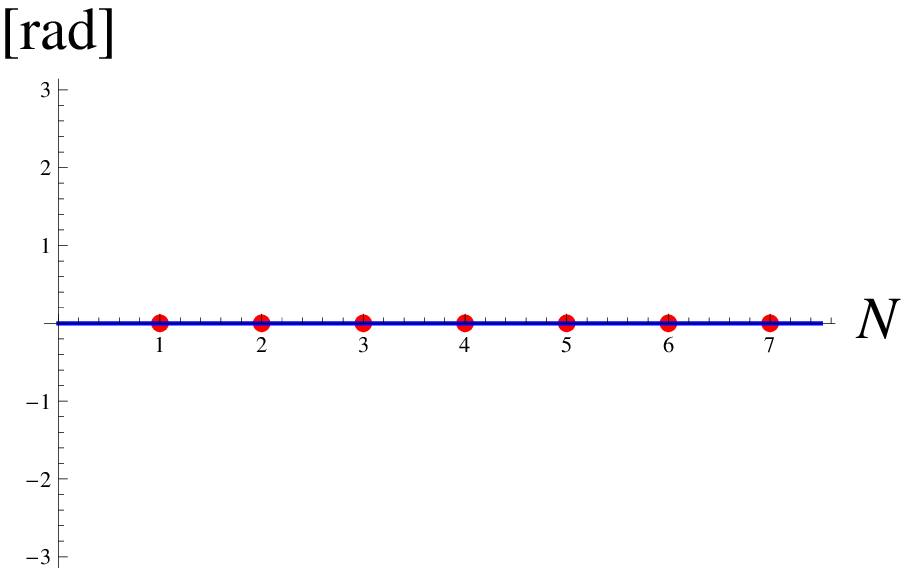}\includegraphics[scale=0.6]{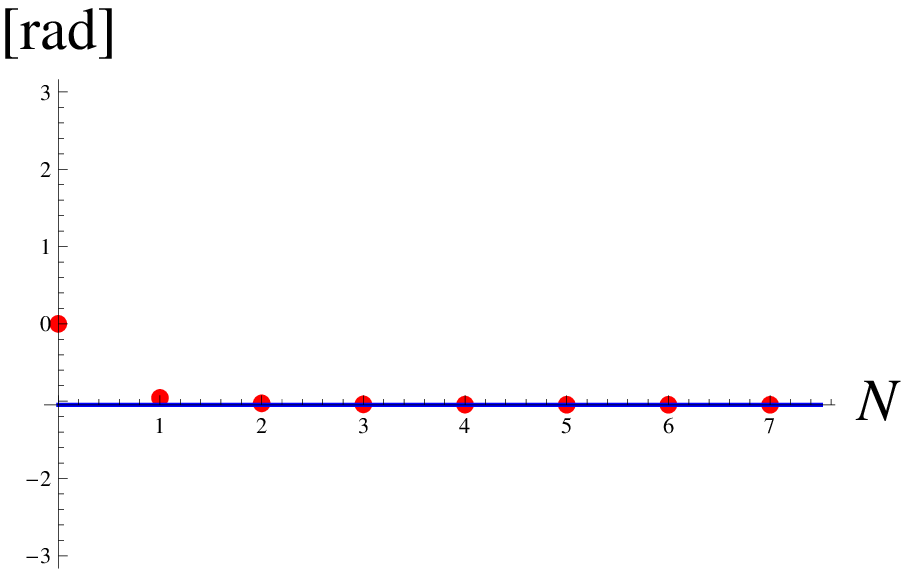}\includegraphics[scale=0.6]{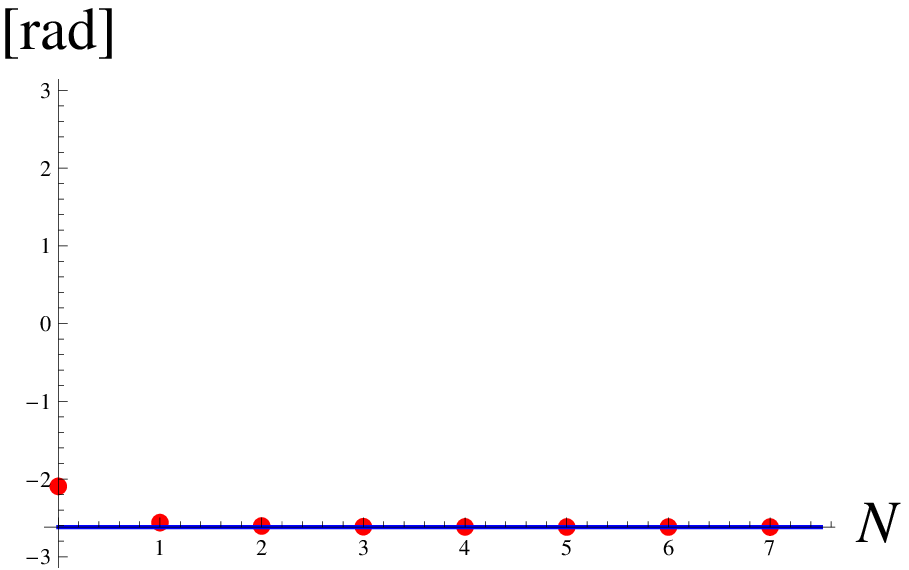}\\
$M=0$\qquad\qquad\qquad\qquad\qquad$M=1$\qquad\qquad\qquad\qquad\qquad$M=2$
\end{center}
\caption{The phases of the two-point function $\langle s_{\Box}\bar s_{\Box}\rangle_{k=6}(N,N+M)$ (red dots) and the phases $\Theta^{\Box,\Box}_{k=6,M}=\theta_{k=6,M}+2\theta_{k=6,M}^{\Box}$ (blue lines) for $M=0$ (left), $M=1$ (center), $M=2$ (right).
The phases are approaching to $\Theta^{\Box,\Box}_{k=6,M}$ in the large $N$ limit for $M=1,2$ and the phase is identically vanishing for $M=0$.
}
\label{BBphase}
\end{figure}

The phase factor of the two-point function is more complicated.
Nevertheless, after plotting the phase, we have found that, in the large $N$ limit, the phase of the two-point function is exponentially approaching to the sum of those of the two one-point functions with separated insertions, 
\begin{align}
\frac{\langle s_Y\bar s_Z\rangle_{k}(N,N+M)}{|\langle s_Y\bar z_Z\rangle_{k}(N,N+M)|}
\to e^{i\Theta^{Y,Z}_{k,M}},\quad
\Theta^{Y,Z}_{k,M}=\theta_{k,M}+\theta_{k,M}^Y+\theta_{k,M}^Z.
\label{twophase}
\end{align}
As an example, in figure \ref{BBphase} we plot the phases of the two-point function $\langle s_{\yng(1)}\bar s_{\yng(1)}\rangle_{k=6}(N,N+M)$ and show how the phases are approaching to $\Theta^{\yng(1),\yng(1)}_{k=6,M}=\theta_{k=6,M}+2\theta_{k=6,M}^{\yng(1)}$ for $M=0,1,2$.

It was known \cite{WCS} that the phase factor is interpreted as the framing factor in the field-theoretical viewpoint.
Here $\theta_{k,M}$ is the framing factor of the manifold while $\theta_{k,M}^Y$ is the framing factor of the Wilson loop.
Hereafter we often consider the two-point function with this framing factor removed
\begin{align}
e^{-i\Theta_{k,M}^{Y,Z}}\langle s_Y\bar s_Z\rangle_{k}(N,N+M).
\end{align}

\subsection{Perturbative part}\label{pertpart}

It turns out that as in the case of the one-point function the result of the two-point function is summarized cleanly in the grand canonical ensemble.
To present the result we define the chemical potential $\mu$ from the fugacity $z$ by $\mu=\log z$ and study the perturbation theory in large $\mu$ in this subsection.

We first conjecture that, for the one-point function of the half-BPS Wilson loop operator in an arbitrary representation, the perturbative part is given by
\begin{align}
\frac{\langle s_Y\rangle^\text{GC}_{k,M}(z)}{\langle 1\rangle^\text{GC}_{k,M}(z)}
\bigg|^\text{pert}
=\frac{e^{i\theta^Y_{k,M}}e^{\frac{2|Y|}{k}\mu}}
{\prod_{(i,j)\in Y}2\sin\frac{2\pi h(i,j)}{k}},
\label{1ptpert}
\end{align}
where $h(i,j)$ at $(i,j)\in Y$ is the hook length
\begin{align}
h(i,j)=\lambda_i+\lambda^\text{T}_j-i-j+1.
\end{align}

There are many consistency checks for this expression.
First, this expression is consistent with the numerical analysis in \cite{HaOk}, where it was found that for many one-point functions in the hook representation, the perturbative part is given as
\begin{align}
\frac{\langle s_{(a|l)}\rangle^\text{GC}_{k,M}(z)}{\langle 1\rangle^\text{GC}_{k,M}(z)}
\bigg|^\text{pert}
=\frac{e^{\frac{\pi i}{k}((a-\frac{M}{2})^2-(l+\frac{M}{2})^2)}e^{\frac{2|Y|}{k}\mu}}
{2\sin\frac{2\pi(a+l)}{k}
\prod_{m=1}^{a-\frac{1}{2}}2\sin\frac{2\pi m}{k}
\prod_{n=1}^{l-\frac{1}{2}}2\sin\frac{2\pi n}{k}}.
\label{1ptperthook}
\end{align}
Our perturbative expression \eqref{1ptpert} reduces to \eqref{1ptperthook} for the hook representation.

Secondly, this expression is also consistent with the Giambelli identity proved in \cite{HHMO,MaMo,FM}
\begin{align}
\frac{\langle s_{(a_1,a_2,\cdots,a_R|l_1,l_2,\cdots,l_R)}\rangle^\text{GC}_{k,M}(z)}
{\langle 1\rangle^\text{GC}_{k,M}(z)}
=\det\biggl(\frac{\langle s_{(a_i|l_j)}\rangle^\text{GC}_{k,M}(z)}
{\langle 1\rangle^\text{GC}_{k,M}(z)}\biggr)
_{\begin{subarray}{c}1\le i\le R\\1\le j\le R\end{subarray}},
\label{Giambelliid}
\end{align}
which reduces the expression in an arbitrary representation to that in the hook representation.
In appendix \ref{giambellic} we prove that the perturbative truncation of the one-point function \eqref{1ptpert} satisfies the Giambelli identity.

Thirdly, after assuming the correspondence with the open topological string theory \cite{HHMO} and the identification of the variables (see \eqref{Vhat} later), we also see that the expression of the open topological string free energy gives \eqref{1ptpert} when expanded for various representations $Y$.
The expansion is given in appendix \ref{freeopen}.

For the perturbative part of the two-point function, from the numerical studies, we find that it is the same as the product of the two one-point functions with separated insertions,
\begin{align}
\frac{\langle s_Y\bar s_Z\rangle^\text{GC}_{k,M}(z)}{\langle 1\rangle^\text{GC}_{k,M}(z)}
\bigg|^\text{pert}
=\frac{\langle s_Y\rangle^\text{GC}_{k,M}(z)}{\langle 1\rangle^\text{GC}_{k,M}(z)}
\bigg|^\text{pert}\times
\frac{\langle\bar s_Z\rangle^\text{GC}_{k,M}(z)}{\langle 1\rangle^\text{GC}_{k,M}(z)}
\bigg|^\text{pert}.
\end{align}

\subsection{Conjugate relation}\label{superposerel}

A direct study of the non-perturbative part of the two-point function is more complicated.
Nevertheless, we can separate the two-point function into the real part and the imaginary part and investigate the large $\mu$ expansion in the grand canonical ensemble as in the one-point function.
We have performed this analysis for a few two-point functions.
After seeing in the previous two subsections that the phase factor and the perturbative part of the two-point function split into the product of those of the two one-point functions with separated insertions, we may expect a more direct relation to the one-point functions even for the non-perturbative part.
In our analysis we have found two relations to the one-point functions.
We shall present one in this subsection, the other in the next subsection and discuss their possible interpretations.

The first relation is
\begin{align}
e^{-i\Theta_{k,M}^{Y,Z}}\langle s_Y\bar s_Z\rangle_{k}(N,N+M)
=\bigl[e^{-i\Theta_{k,M}^{Y,Z}}\langle s_Ys_Z\rangle_{k}(N,N+M)\bigr]^*.
\label{superposeid}
\end{align}
Namely, our two-point function with the main phase factor removed is equal to the complex conjugate of the trivial two-point function constructed by the same two characters and can be decomposed into the one-point functions following the Littlewood-Richardson rule.
From our original viewpoint of the definition of the non-trivial two-point function \eqref{two}, there are no signs that we can superpose the two insertions.
Nevertheless, after the long analysis of the residue computations, we have found that actually these two insertions can be superposed to each other and essentially reduced to the one-point functions.
In the following, we shall refer to this relation as the conjugate relation.

There are several ways to express this relation.
For example, if we do not want our result to contain the phase factors, we can express the result as
\begin{align}
\frac{\langle s_Y\bar s_Z\rangle_{k}(N_1,N_2)\cdot\langle 1\rangle_{k}(N_1,N_2)}
{\langle s_Y\rangle_{k}(N_1,N_2)\cdot\langle\bar s_Z\rangle_{k}(N_1,N_2)}
=\biggl[\frac{\langle s_Ys_Z\rangle_{k}(N_1,N_2)\cdot\langle 1\rangle_{k}(N_1,N_2)}
{\langle s_Y\rangle_{k}(N_1,N_2)\cdot\langle s_Z\rangle_{k}(N_1,N_2)}\biggr]^*.
\end{align}
Also, if we expand the two characters by the Littlewood-Richardson rule $s_Ys_Z=\sum_XN_{YZ}^Xs_X$, it is given by
\begin{align}
e^{-2i\Theta_{k,M}^{Y,Z}}\langle s_Y\bar s_Z\rangle_{k}(N,N+M)
=\sum_XN_{YZ}^X
e^{-2i\Theta_{k,M}^{X}}\langle s_X\rangle_{k}(N,N+M).
\label{2ptfrom1pt}
\end{align}

Although this relation is natural from the viewpoint of our studies of the phase factor and the perturbative part, it is a highly non-trivial relation, considering that it continues to be valid for all of the non-perturbative corrections.
In fact, at present we cannot prove this relation from any rigorous arguments.
We have checked this relation for all the cases of $2\le |Y|+|Z|\le 5$ and $k=3,4,6,8,12$ where the integrations are convergent.
Especially we have listed some data for the case of $\langle s_\Box\bar s_\Box\rangle_{k=6}(N,N+M)$ with $M=0,1,2$ in appendix \ref{superpose} to convince the readers of the validity.
We note that, although all the ingredients in the two-point function $\langle s_\Box\bar s_\Box\rangle_{k=6}(N,N+M)$ and the one-point function $\langle s_{\yng(2)}\rangle_{k=6}(N,N+M)$ are convergent for $M=0,1,2$, those in the one-point function $\langle s_{\yng(1,1)}\rangle_{k=6}(N,N+M)$ are convergent only for $M=0,1$.
The ranges of the convergence do not necessarily coincide for the both sides of \eqref{2ptfrom1pt}.

This relation may reflect the topological aspects of the ABJM matrix model.
Although the original ABJM theory is not topological, after applying the localization techniques for the one-point function of the half-BPS Wilson loop, the result is known to reduce to the matrix model and relate to the topological string theory.
Our definition of the two-point function is not obtained from the localization techniques and is not related to topological theories at the first sight.
However, since this definition respects all of the symmetries the one-point function has, it may still relate to the topological string theory.
Therefore, we expect that the relation we have found is another sign of the deep relation between the ABJM matrix model and the topological string theory, though the full interpretation is unclear to us.

\subsection{Descent relation}\label{interfererel}

As we have explained in section \ref{secphase}, in the large $N$ limit, the phase of the two-point function $\langle s_Y\bar s_Z\rangle_k(N,N+M)$ is approaching to $\Theta^{Y,Z}_{k,M}$ exponentially.
This implies that the imaginary part of the two-point function with the main phase factor removed, $\im e^{-i\Theta^{Y,Z}_{k,M}}\langle s_Y\bar s_Z\rangle_k(N,N+M)$, has a simpler structure.
In this subsection, we observe an interesting relation for the imaginary part of the two-point function.

In the study of the simplest two-point function $\langle s_\Box\bar s_\Box\rangle_k(N,N+M)$ with both insertions in the fundamental representation $\Box$, we find that, after removing the main phase, the imaginary part of the two-point function simply reduces to the partition function
\begin{align}
\im\Bigl[e^{-i\Theta_{k,M}^{\Box,\Box}}\langle s_\Box\bar s_\Box\rangle_k(N,N+M)\Bigr]
=\biggl(\sin\frac{2\pi M}{k}\biggr)e^{-i\Theta_{k,M}^{\varnothing}}\langle 1\rangle_k(N,N+M).
\end{align}
It turns out that this relation is a special case of a more general relation between the imaginary part of the two-point function and the one-point functions.
Our conjecture is that there exists a set of Laurent polynomials $Q_{YZ}^X$ of $q=e^{-\frac{4\pi i}{k}}$ depending only on the Young diagrams so that the relation
\begin{align}
\im\Bigl[e^{-i\Theta_{k,M}^{Y,Z}}\langle s_Y\bar s_Z\rangle_{k}(N,N+M)\Bigr]
=\im\Bigl[e^{-i\Theta_{k,M}^{Y,Z}}\sum_{|X|=|Y|+|Z|-2}Q_{YZ}^X\langle s_X\rangle_{k}(N,N+M)\Bigr],
\label{imaginary}
\end{align}
holds where the sum is taken over the Young diagrams $X$ whose box number $|X|$ is less than $|Y|+|Z|$ by $2$.
Later we will denote this relation simply by
\begin{align}
\langle s_Y\bar s_Z\rangle\sim\sum_XQ_{YZ}^X\langle s_X\rangle
\quad\mod e^{i\Theta^{Y,Z}}{\mathbb R},
\label{generalinterfere}
\end{align}
or, when there is no confusion, we often drop mod $e^{i\Theta^{Y,Z}}{\mathbb R}$ since the phase is clear for the two-point function.

For the case of $Z=\Box$, we have explicitly found a simple set of the coefficients $Q_{Y\Box}^X$ for the relation \eqref{imaginary}.
Namely, $Q^X_{Y\Box}$ is non-vanishing only for $X=Y_\bullet$ with $Y_\bullet$ denoting the Young diagram with one box removed from $Y$ without affecting the rule of the Young diagram and $Q_{Y\Box}^{Y_\bullet}=1$.
In other words, the relation simplifies to
\begin{align}
\langle s_Y\bar s_\Box\rangle
\sim\sum_{Y_\bullet}\langle s_{Y_\bullet}\rangle.
\label{generalbox}
\end{align}
Especially when $Y=(a|l)$ is the hook representation, the relation is
\begin{align}
\langle s_{(a|l)}\bar s_\Box\rangle
\sim\langle s_{(a|l-1)}\rangle+\langle s_{(a-1|l)}\rangle.
\label{hookbox}
\end{align}
Also, for the case of $Y=(a|\frac{1}{2})$ and $Z=\yng(2)$, the relation is
\begin{align}
\langle s_{(a|\frac{1}{2})}\bar s_{\yng(2)}\rangle
\sim q^{a-\frac{1}{2}}\langle s_{(a|\frac{1}{2})}\rangle
+q^{-1}\langle s_{(a-1|\frac{3}{2})}\rangle,
\label{hooksym}
\end{align}
while for the case of $Y=(\frac{1}{2}|l)$ and $Z=(a|\frac{1}{2})$, the relation is
\begin{align}
\langle s_{(\frac{1}{2}|l)}\bar s_{(a|\frac{1}{2})}\rangle
\sim\langle s_{(a|l-1)}\rangle+\langle s_{(a-1|l)}\rangle.
\label{antisymsym}
\end{align}
In fact \eqref{hookbox} and \eqref{antisymsym} are equivalent if we assume the conjugate relation \eqref{superposeid} in the previous subsection.
We list the first few concrete relations in appendix \ref{interference}.
For the cases when the total box number is less than five, the relations are always special cases of our list in \eqref{generalbox}, \eqref{hooksym}, \eqref{antisymsym}, though for the cases with more boxes there appear relations which do not belong to the cases we have discussed.
Nevertheless, our conjecture is that they still satisfy the general expression \eqref{imaginary} with a suitable choice of the Laurent polynomials $Q_{YZ}^X$.

Note that we do not claim that the set of the coefficients $Q_{YZ}^X$ is unique due to the non-trivial relations for the imaginary part among the one-point functions with the same box number
\begin{align}
\im\Bigl[e^{-i\Theta^{(n)}_{k,M}}\sum_{|X|=n}Q^X\langle s_X\rangle_k(N,N+M)\Bigr]=0,
\label{ambiX}
\end{align}
with
\begin{align}
\Theta^{(n)}_{k,M}=\theta_{k,M}-(n+2)\frac{\pi M}{k},
\end{align}
or, in our abbreviated notation,
\begin{align}
\sum_{|X|=n}Q^X\langle s_X\rangle\sim 0\quad\mod e^{i\Theta^{(n)}}{\mathbb R}.
\label{gen1pt}
\end{align}
The set of the coefficients $Q_{YZ}^X$ is determined only up to these ambiguities and we have chosen one representative set to express the relation.
Nevertheless, we stress that, once $Q_{YZ}^X$ is chosen up to the ambiguities, the same set of $Q_{YZ}^X$ is valid for any $N$ and $M$.

For example, for the case of the one-point functions with three boxes, we find a non-trivial relation for the one-point functions
\begin{align}
\im\Bigl[
e^{-i\Theta^{(3)}_{k,M}}
(q\langle s_{\yng(3)}\rangle_k(N,N+M)
-\langle s_{\yng(2,1)}\rangle_k(N,N+M)
+q^{-1}\langle s_{\yng(1,1,1)}\rangle_k(N,N+M))\Bigr]=0.
\label{onepointthree}
\end{align}
Note that we are free to multiply the relation \eqref{onepointthree} by
\begin{align}
[n]_q=\frac{q^{\frac{n}{2}}-q^{-\frac{n}{2}}}{q^{\frac{1}{2}}-q^{-\frac{1}{2}}},
\label{qnumber}
\end{align}
since $[n]_q\in{\mathbb R}$, which is sometimes necessary for discussing the representative choice of $Q^X_{YZ}$.
In the notation \eqref{gen1pt}, the relation for the one-point functions can be expressed as
\begin{align}
q\langle s_{\yng(3)}\rangle-\langle s_{\yng(2,1)}\rangle+q^{-1}\langle s_{\yng(1,1,1)}\rangle
\sim 0\quad\mod e^{i\Theta^{(3)}}{\mathbb R}.
\label{onerel3}
\end{align}
For the case of the one-point functions with more than three boxes, there are more similar relations.
We list the first few ambiguities in appendix \ref{ambiguity}.
Our descent relations in \eqref{generalbox}, \eqref{hooksym}, \eqref{antisymsym} and appendix \ref{interference} are given up to these ambiguities.

We have many checks for the relation \eqref{imaginary} (and \eqref{ambiX} as well).
Since we have computed the exact values of the two-point function up to a certain rank $N=N_\text{max}$, we can substitute the values to the relations to check the validity.
Also in appendix \ref{lowest} we have computed the non-vanishing two-point function of the lowest rank (or the lowest component in the grand canonical ensemble) and we check the relation \eqref{imaginary} for the lowest component in appendix \ref{lowcheck}.
Anyway, our conjecture passes all of the consistency checks.

Since the lowest component is determined for any values of $M$, we can proceed one step further by asking which set of the Laurent polynomials $Q_{YZ}^X$ satisfies the relation for the lowest component.
Surprisingly, we find that actually the lowest component gives a large enough number of the constraints so that as long as these constraints are satisfied the relation on the imaginary part also holds for the higher components.
In fact this is true for all of the relations given in appendix \ref{interference}.

The interpretation of the relation \eqref{imaginary} is again obscure.
Since we are considering the imaginary part of the two-point function, this may relate to the interference between the two insertions.
One may also expect that the relation is related to the orientifold projection \cite{MS1,Hosp,Oosp,MS2,MN5}.
On one hand, the restriction to the imaginary part can be regarded as a projection.
On the other hand, the resulting condition for the non-vanishing values of $Q_{YZ}^X$, $|X|=|Y|+|Z|-2$, is reminiscent of the reduction of the unitary groups to the orthogonal groups or the symplectic groups (due to the invariant tensor $\delta_{ab}$ and $J_{ab}$ respectively).
An alternative attempt for the interpretation is given in appendix \ref{heisenberg}, where the reduction by two boxes is interpreted by the derivative, in the analogy of the symplectic structure of the Heisenberg algebra $[\widehat q,\widehat p]=i\hbar$, which reduces the total number of the coordinate/momentum operators by two.
Since the relation reduces the total box number, we shall refer to this relation as the descent relation.

\subsection{Implications to one-point functions}\label{oneptrel}

In the previous two subsections, we have found two relations among the two-point functions and the one-point functions.
The conjugate relation reduces our non-trivial two-point function to the trivial two-point function which can be reexpanded into the one-point functions by the Littlewood-Richardson rule.
The descent relation reduces the two-point function to the one-point functions more directly by concentrating on the imaginary part.
Although originally the relations are naturally given in terms of the two-point function, after eliminating the two-point function, we find a relation purely among the one-point functions with the box numbers differing by two.
For example, for the case of two boxes, the relation is
\begin{align}
\sin\frac{2\pi}{k}|\langle s_{\yng(2)}\rangle_k(N_1,N_2)|
-\sin\frac{2\pi}{k}|\langle s_{\yng(1,1)}\rangle_k(N_1,N_2)|
+\sin\frac{2\pi M}{k}|\langle 1\rangle_k(N_1,N_2)|=0.
\end{align}
We shall abbreviate this relation as
\begin{align}
\sigma_1|\langle s_{\yng(2)}\rangle|-\sigma_1|\langle s_{\yng(1,1)}\rangle|
+\sigma_M|\langle 1\rangle|=0,
\end{align}
by introducing $\sigma_n=\sin(\frac{2\pi}{k}n)$. 
We can alternatively write down the expression using the $q$-number \eqref{qnumber} with $[n]_q=\sigma_n/\sigma_1$.
Similar relations also hold for more boxes.
We summarize the relations for the box number up to five in appendix \ref{onept}.
Beyond four boxes the relations are subject to the ambiguities \eqref{onepointthree} as we have discussed.
These relations are given purely with the one-point functions and some of them reproduce the relations known previously.
We could have pointed out these relations without introducing the two-point function.
However, with the two simple relations given in the previous two subsections, we find it more natural to discuss in a larger framework with the two-point function.

\section{Topological strings}\label{onepoint}

In the previous section we have defined the non-trivial two-point function, studied them carefully and found that the two-point function relates to the one-point function directly.
From these studies we are convinced of the importance of the two-point function we have defined, since it provides a natural framework to unify many aspects of the one-point function.  
Here on occation of various numerical data, let us revisit the one-point function and uncover some fine structure of them.
Among others, we find that the Gopakumar-Vafa invariants are asymmetric in the exchange of the two degrees, which is not very common in our experience.
We first briefly explain the correspondence between the matrix model and the topological string theory before going into the analysis of the Gopakumar-Vafa invariants.
Although most of the correspondences are well-known, we try to shed some light by presenting the cases of closed strings and open strings in a parallel manner.

\subsection{Partition function and closed topological strings}

From the definition of the matrix model in the grand canonical ensemble \eqref{gc}, it is not difficult to observe that the function $\langle s_Y\bar s_{Y'}\rangle^\text{GC}_{k,M}(e^\mu)$ is invariant under the shift of $\mu$ by $2\pi i$.
For this reason, we define the {\it reduced} correlation function by
\begin{align}
\langle s_Y\bar s_{Y'}\rangle^\text{GC}_{k,M}(e^\mu)
=\sum_{n=-\infty}^\infty
\llangle s_Y\bar s_{Y'}\rrangle^\text{GC}_{k,M}(\mu+2\pi in).
\end{align}
Then it was known that, if we further redefine the chemical potential $\mu$ into $\mu_\text{eff}$, the reduced grand canonical partition function can be described by the free energy of closed topological strings \cite{HMMO}
\begin{align}
\llangle 1\rrangle^\text{GC}_{k,M}(\mu)=e^{F^\text{cl}({\bm T})}.
\label{Fcl}
\end{align}
Here, besides the perturbative part $F^\text{pert}({\bm T})$, the free energy
\begin{align}
F^\text{cl}({\bm T})=F^\text{pert}({\bm T})+F^\text{WS}({\bm T})+F^\text{MB}({\bm T}),
\end{align}
is separated into the worldsheet instanton part $F^\text{WS}({\bm T})$ and the membrane instanton part $F^\text{MB}({\bm T})$ given by $(s_\text{L/R}=2j_\text{L/R}+1)$
\begin{align}
F^\text{WS}({\bm T})
&=\sum_{\bm d}\sum_{j_\text{L},j_\text{R}}N^{\bm d}_{j_\text{L},j_\text{R}}
\sum_{n=1}^\infty\frac{(-1)^{(s_\text{L}+s_\text{R}-1)n}s_\text{R}\sin 2\pi g_\text{s}ns_\text{L}}
{n(2\sin\pi g_\text{s}n)^2\sin 2\pi g_\text{s}n}e^{-n{\bm d}\cdot{\bm T}},\nonumber\\
F^\text{MB}({\bm T})
&=\sum_{\bm d}\sum_{j_\text{L},j_\text{R}}N^{\bm d}_{j_\text{L},j_\text{R}}
\sum_{n=1}^\infty\frac{\partial}{\partial g_\text{s}}
\biggl[g_\text{s}
\frac{-\sin\frac{\pi n}{g_\text{s}}s_\text{L}\sin\frac{\pi n}{g_\text{s}}s_\text{R}}
{4\pi n^2(\sin\frac{\pi n}{g_\text{s}})^3}e^{-n\frac{{\bm d}\cdot{\bm T}}{g_\text{s}}}\biggr],
\end{align}
with the identification
\begin{align}
T_\pm=\frac{4\mu_\text{eff}}{k}\pm\pi i\biggl(1-\frac{2M}{k}\biggr),\quad
g_\text{s}=\frac{2}{k}.
\label{tgs}
\end{align}
For integral $k$ and $M$ the effective chemical potential is explicitly given by
\begin{align}
\mu_\text{eff}=\begin{cases}\displaystyle\mu-2(-1)^{\frac{k}{2}-M}e^{-2\mu}
{}_4F_3\biggl(1,1,\frac{3}{2},\frac{3}{2};2,2,2;16(-1)^{\frac{k}{2}-M}e^{-2\mu}\biggr),
&k:\text{even},\\
\displaystyle\mu+e^{-4\mu}
{}_4F_3\biggl(1,1,\frac{3}{2},\frac{3}{2};2,2,2;-16e^{-4\mu}\biggr),
&k:\text{odd}.
\end{cases}
\label{integermu}
\end{align}

When the BPS index $N^{\bm d}_{j_\text{L},j_\text{R}}$ is non-vanishing only for $s_\text{L}+s_\text{R}-1\equiv 0$ mod $2$, if we ignore the non-perturbative membrane instanton part $F^\text{MB}({\bm T})$, we can rewrite the worldsheet instanton part $F^\text{WS}({\bm T})$ as
\begin{align}
F^\text{WS}({\bm T})=\sum_{\bm d}\sum_{g=0}^\infty\sum_{n=1}^\infty
n^{\bm d}_g\frac{(2i\sin\pi g_\text{s}n)^{2g-2}}{n}e^{-n{\bm d}\cdot{\bm T}},
\end{align}
where we have introduced the Gopakumar-Vafa invariants $n^{\bm d}_g$ by
\begin{align}
\sum_{j_\text{L},j_\text{R}}N^{\bm d}_{j_\text{L},j_\text{R}}
\frac{s_\text{R}\sin 2\pi g_\text{s}ns_\text{L}}{(2\sin\pi g_\text{s}n)^2\sin 2\pi g_\text{s}n}
=\sum_{g=0}^\infty n^{\bm d}_g(2i\sin\pi g_\text{s}n)^{2g-2}. 
\end{align}
The free energy of closed topological strings was known to enjoy the multi-covering property.
Namely, if we define the multi-covering component as $(Q_\pm=e^{-T_\pm})$
\begin{align}
A_{k}({\bm Q})=\sum_{\bm d}\sum_{g=0}^\infty n^{\bm d}_g
\biggl(2i\sin\frac{2\pi}{k}\biggr)^{2g-2}
{\bm Q}^{\bm d},
\end{align}
or more explicitly
\begin{align}
A_{k,M}(Q)=\sum_{\bm d}\sum_{g=0}^\infty n^{\bm d}_g
\biggl(2i\sin\frac{2\pi}{k}\biggr)^{2g-2}
e^{(d_+-d_-)\frac{2\pi iM}{k}}Q^{d_++d_-},
\end{align}
after introducing $Q=-e^{-\frac{4\mu_\text{eff}}{k}}$, the free energy can be expressed as
\begin{align}
F^\text{WS}({\bm T})=\sum_{n=1}^\infty\frac{1}{n}A_{\frac{k}{n},M}(Q^n).
\end{align}
Higher powers of $Q$ in $F^\text{WS}({\bm T})$ consist both of the Gopakumar-Vafa invariants of higher degrees and the Gopakumar-Vafa invariants of lower degrees.
Physically, by regrading $n$ as the winding number, this is interpreted that both genuine states of multiple degrees without windings and states of degree one wound multiply contribute as the same effects.

\subsection{One-point functions and open topological strings}

The normalized one-point function in the grand canonical ensemble is given by the free energy of open topological strings
\begin{align}
\sum_Y\frac{\llangle s_Y\rrangle^\text{GC}_{k,M}(\mu)}
{\llangle 1\rrangle^\text{GC}_{k,M}(\mu)}\tr_Y(V)
=e^{F^\text{op}({\bm T},\widehat V)},
\label{mattop}
\end{align}
where the free energy of open topological strings is given by
\begin{align}
F^\text{op}({\bm T},\widehat V)
=\sum_{\bm d}\sum_{g=0}^\infty\sum_{h=1}^\infty\sum_{\bm\ell}\sum_{n=1}^\infty
n^{\bm d,\bm\ell}_{g}\frac{(2i\sin\pi g_\text{s}n)^{2g-2}}{n}
\frac{1}{h!}
\prod_{j=1}^h\biggl(\frac{2i\sin\pi g_\text{s}n\ell_j}{\ell_j}\tr\widehat V^{n\ell_j}\biggr)
e^{-n{\bm d}\cdot{\bm T}},
\label{openfree}
\end{align}
with the identification
\begin{align}
\widehat V=Q_+^{-\frac{1}{2}}V,
\label{Vhat}
\end{align}
along with those in the partition function $Q_\pm=e^{-T_\pm}$, \eqref{tgs}, \eqref{integermu}.
Note that the relation in \cite{GKM,HHMO}
\begin{align}
\frac{\llangle e^{\sum_{n=1}^\infty
\frac{1}{n}\Str U^n\tr V^n}\rrangle^\text{GC}_{k,M}(\mu)}
{\llangle 1\rrangle^\text{GC}_{k,M}(\mu)}
=e^{F^\text{op}({\bm T},\widehat V)},
\end{align}
with $U=\diag(e^{\mu_1},\cdots,e^{\mu_N}|{-e^{\nu_1}},\cdots,-e^{\nu_{N+M}})$ is rewritten into \eqref{mattop} with the help of the orthogonal relation
\begin{align}
\frac{\prod_{j,k}(1+y_jz_k)}{\prod_{i,k}(1-x_iz_k)}=\sum_Ys_Y(x|y)s_Y(z),
\end{align}
where $s_Y(x|y)$ is the super Schur polynomial while $s_Y(z)$ is the ordinary Schur polynomial.

The identification \eqref{Vhat} without the fractional branes $M=0$ was pointed out in \cite{GKM} with $\widehat V=Q^{-\frac{1}{2}}V$.
Here we generalize the identification to the cases with $M\ne 0$ by replacing $Q$ with only one of the Kahler parameters $Q_+=e^{-T_+}$.
In appendix \ref{freeopen}, by carefully studying the phase factor as well, we check that the identification \eqref{Vhat} is consistent with the correspondence between the open topological string theory \eqref{mattop} and the perturbative part of the one-point function in the ABJM matrix model \eqref{1ptpert} in section \ref{pertpart}.

As in the case of closed topological strings, we can express the free energy as
\begin{align}
F^\text{op}({\bm T},\widehat V)=\sum_{h=1}^\infty\sum_{\bm\ell}\sum_{n=1}^\infty
\frac{1}{n}A^{\bm\ell}_{\frac{k}{n},M}(Q^n)\frac{1}{h!}\prod_{j=1}^h\frac{\tr V^{nl_j}}{l_j},
\label{multicoverF}
\end{align}
if we define the multi-covering component suitably.
We first define the multi-covering component as
\begin{align}
A_k^{\bm\ell}(Q_+,Q_-)=\sum_{\bm d}\sum_{g=0}^\infty n^{\bm d,\bm\ell}_g
\biggl(2i\sin\frac{2\pi}{k}\biggr)^{2g-2}\biggl(\prod_{j=1}^h2i\sin\frac{2\pi l_j}{k}\biggr)
(Q_+^{-\frac{1}{2}})^{|{\bm\ell}|-2d_+}(Q_-^{-\frac{1}{2}})^{-2d_-},
\end{align}
with $(Q_+^{-\frac{1}{2}},Q_-^{-\frac{1}{2}})=(ie^{\frac{2\mu_\text{eff}}{k}}e^{-\frac{\pi iM}{k}},-ie^{\frac{2\mu_\text{eff}}{k}}e^{\frac{\pi iM}{k}})$, and observe that we can safely change the sign of $Q_-^{-\frac{1}{2}}=-ie^{\frac{2\mu_\text{eff}}{k}}e^{\frac{\pi iM}{k}}$ into $Q_-^{-\frac{1}{2}}=ie^{\frac{2\mu_\text{eff}}{k}}e^{\frac{\pi iM}{k}}$ since only even powers appear.
This means that we can rewrite the multi-covering component into
\begin{align}
A_{k,M}^{\bm\ell}(Q)=\sum_{\bm d}\sum_{g=0}^\infty n^{\bm d,\bm\ell}_g
\biggl(2i\sin\frac{2\pi}{k}\biggr)^{2g-2}\biggl(\prod_{j=1}^h2i\sin\frac{2\pi l_j}{k}\biggr)
(e^{-\frac{\pi i M}{k}})^{|{\bm\ell}|-2(d_+-d_-)}
(Q^{-\frac{1}{2}})^{|{\bm\ell}|-2(d_++d_-)},
\end{align}
with $Q^{-\frac{1}{2}}=ie^{\frac{2\mu_\text{eff}}{k}}$.

Note that, although the free energy of open topological strings is clearly given in the power sum basis, the one-point function in the ABJM matrix model is more naturally given in the basis of the Schur function with the universal phase \eqref{1ptphase}.
To relate these two theories, as the first step, let us see how the multi-covering structure is given in the basis of the Schur function.
For this purpose we expand
\begin{align}
F^\text{op}({\bm T},\widehat V)=A^{(1)}_{1}\tr V
+\frac{A^{(1)}_{2}}{2}\tr V^2+\cdots
+A^{(1,1)}_{1}\frac{(\tr V)^2}{2}+\cdots+A^{(2)}_{1}\frac{\tr V^2}{2}+\cdots,
\end{align}
with the abbreviation $A^{\bm\ell}_{n}=A^{\bm\ell}_{\frac{k}{n},M}(Q^n)$ for \eqref{multicoverF}.
Then, we find that the exponentiation can be expanded by
\begin{align}
e^{F^\text{op}({\bm T},\widehat V)}=\sum_Y\widetilde A^Y\tr_Y(V),
\end{align}
where each coefficient is
\begin{align}
&\widetilde A^{\yng(1)}=A^{(1)}_{1},\nonumber\\
&\widetilde A^{\yng(2)}=\biggl[\frac{1}{2}A^{(1,1)}_{1}+\frac{1}{2}A^{(2)}_{1}\biggr]
+\biggl[\frac{1}{2}(A^{(1)}_{1})^2+\frac{1}{2}A^{(1)}_{2}\biggr],\nonumber\\
&\widetilde A^{\yng(1,1)}=\biggl[\frac{1}{2}A^{(1,1)}_{1}-\frac{1}{2}A^{(2)}_{1}\biggr]
+\biggl[\frac{1}{2}(A^{(1)}_{1})^2-\frac{1}{2}A^{(1)}_{2}\biggr].
\end{align}
For $\widetilde A^{\yng(2)}$ and $\widetilde A^{\yng(1,1)}$ it is natural to interpret the terms in the first bracket as the genuine states without windings and the terms in the second bracket as the states coming from the double windings of $\widetilde A^{\yng(1)}$.
This computation continues to higher degrees.

Comparing the proposals for the partition function and the one-point function from the viewpoint of the tau function of the integrable system, we find a close similarity.
It was observed in \cite{BGT} that the partition function of the spectral curve (generalizing the ABJM matrix model) corresponds to the tau function of the $q$-Painleve equation and it was proved in \cite{MaMo,FM} that the normalized one-point function in the ABJM matrix model satisfies the Giambelli and Jacobi-Trudi identities, which are shared with the expansion coefficients of the tau function of the soliton theory \cite{Sato,MJD,AKLTZ}.
These two proposals seem compatible with each other by regarding $e^{F^\text{cl}}$ and $e^{F^\text{op}}$ as the ``closed'' and ``open'' tau functions respectively.
Also if we combine the proposals \eqref{Fcl} and \eqref{mattop}, we find
\begin{align}
\sum_Y\llangle s_Y\rrangle^\text{GC}_{k,M}(\mu)\tr_Y(V)
=e^{F^\text{cl}({\bm T})+F^\text{op}({\bm T},\widehat V)}.
\end{align}
This relation combining closed topological strings and open topological strings may make the open-closed duality between the ABJM matrix model \cite{HaOk,KM} and the topological strings \cite{GO1,GO2} clearer.

\subsection{BPS indices}

\begin{table}[ht!]
\begin{center}
\begin{tabular}{|c||c|c|c|c|c|c|}
\hline
&$0$&$1$&$2$&$3$&$4$&$5$\\
\hline\hline
$0$&$1$&$1$&$0$&$0$&$0$&$0$\\
\hline
$1$&$1$&$3$&$5$&$7$&$9$\\
\cline{1-6}
$2$&$0$&$5$&$35$&$135$\\
\cline{1-5}
$3$&$0$&$7$&$135$\\
\cline{1-4}
$4$&$0$&$9$\\
\cline{1-3}
$5$&$0$\\
\cline{1-2}
\end{tabular}
\begin{tabular}{|c||c|c|c|c|c|c|}
\hline
&$0$&$1$&$2$&$3$&$4$&$5$\\
\hline\hline
$0$&$0$&$0$&$0$&$0$&$0$&$0$\\
\hline
$1$&$0$&$0$&$0$&$0$&$0$\\
\cline{1-6}
$2$&$0$&$0$&$8$&$72$\\
\cline{1-5}
$3$&$0$&$0$&$72$\\
\cline{1-4}
$4$&$0$&$0$\\
\cline{1-3}
$5$&$0$\\
\cline{1-2}
\end{tabular}
\begin{tabular}{|c||c|c|c|c|c|c|}
\hline
&$0$&$1$&$2$&$3$&$4$&$5$\\
\hline\hline
$0$&$0$&$0$&$0$&$0$&$0$&$0$\\
\hline
$1$&$0$&$0$&$0$&$0$&$0$\\
\cline{1-6}
$2$&$0$&$0$&$0$&$11$\\
\cline{1-5}
$3$&$0$&$0$&$11$\\
\cline{1-4}
$4$&$0$&$0$\\
\cline{1-3}
$5$&$0$\\
\cline{1-2}
\end{tabular}\\
$n^{{\bm d},{\bm\ell}=(1)}_{g=0}$\hspace{36mm}
$n^{{\bm d},{\bm\ell}=(1)}_{g=1}$\hspace{36mm}
$n^{{\bm d},{\bm\ell}=(1)}_{g=2}$
\caption{The Gopakumar-Vafa invariants $n^{{\bm d},{\bm\ell}}_g$ for $\bm\ell=(1)$.
Each column and each row denote the specific values of $d_+$ and $d_-$ respectively.}
\label{Y1}
\end{center}
\end{table}

After the review of the correspondence, let us turn to the study of the Gopakumar-Vafa invariants in the open topological string theory.
Although in \cite{HHMO} it was found that the one-point function of the half-BPS Wilson loop is described by the diagonal BPS indices identified in \cite{GKM}, it was not known how the diagonal BPS indices are split.
Especially, since some of the diagonal BPS indices given in \cite{GKM} are odd which can only be decomposed by the degree difference asymmetrically, it is interesting to see how the diagonal BPS indices are split.
With the abundant numerical data, here we read off the non-perturbative one-point function in the grand canonical ensemble in appendix \ref{np1pt} and identify the split for the first few BPS indices.

\begin{table}[ht!]
\begin{center}
\begin{tabular}{|c||c|c|c|c|c|c|}
\hline
&$0$&$1$&$2$&$3$&$4$&$5$\\
\hline\hline
$0$&$0$&$0$&$0$&$0$&$0$&$0$\\
\hline
$1$&$1$&$2$&$4$&$6$&$8$\\
\cline{1-6}
$2$&$0$&$4$&$36$&$160$\\
\cline{1-5}
$3$&$0$&$6$&$160$\\
\cline{1-4}
$4$&$0$&$8$\\
\cline{1-3}
$5$&$0$\\
\cline{1-2}
\end{tabular}
\begin{tabular}{|c||c|c|c|c|c|c|}
\hline
&$0$&$1$&$2$&$3$&$4$&$5$\\
\hline\hline
$0$&$0$&$0$&$0$&$0$&$0$&$0$\\
\hline
$1$&$0$&$0$&$0$&$0$&$0$\\
\cline{1-6}
$2$&$0$&$0$&$7$&$74$\\
\cline{1-5}
$3$&$0$&$0$&$74$\\
\cline{1-4}
$4$&$0$&$0$\\
\cline{1-3}
$5$&$0$\\
\cline{1-2}
\end{tabular}
\begin{tabular}{|c||c|c|c|c|c|c|}
\hline
&$0$&$1$&$2$&$3$&$4$&$5$\\
\hline\hline
$0$&$0$&$0$&$0$&$0$&$0$&$0$\\
\hline
$1$&$0$&$0$&$0$&$0$&$0$\\
\cline{1-6}
$2$&$0$&$0$&$0$&$10$\\
\cline{1-5}
$3$&$0$&$0$&$10$\\
\cline{1-4}
$4$&$0$&$0$\\
\cline{1-3}
$5$&$0$\\
\cline{1-2}
\end{tabular}\\
$n^{{\bm d},{\bm\ell}=(1,1)}_{g=0}$\hspace{36mm}
$n^{{\bm d},{\bm\ell}=(1,1)}_{g=1}$\hspace{36mm}
$n^{{\bm d},{\bm\ell}=(1,1)}_{g=2}$
\caption{The Gopakumar-Vafa invariants $n^{{\bm d},{\bm\ell}}_g$ for $\bm\ell=(1,1)$.
The asymmetry of the Gopakumar-Vafa invariants in the degrees appear in $n^{(d_+,d_-)=(0,1),{\bm\ell}=(1,1)}_{g=0}=1$.}
\label{Y11}
\end{center}
\end{table}

\begin{table}[ht!]
\begin{center}
\begin{tabular}{|c||c|c|c|c|c|c|}
\hline
&$0$&$1$&$2$&$3$&$4$&$5$\\
\hline\hline
$0$&$0$&$0$&$0$&$0$&$0$&$0$\\
\hline
$1$&$1$&$2$&$4$&$6$&$8$\\
\cline{1-6}
$2$&$0$&$4$&$24$&$96$\\
\cline{1-5}
$3$&$0$&$6$&$96$\\
\cline{1-4}
$4$&$0$&$8$\\
\cline{1-3}
$5$&$0$\\
\cline{1-2}
\end{tabular}
\begin{tabular}{|c||c|c|c|c|c|c|}
\hline
&$0$&$1$&$2$&$3$&$4$&$5$\\
\hline\hline
$0$&$0$&$0$&$0$&$0$&$0$&$0$\\
\hline
$1$&$0$&$0$&$0$&$0$&$0$\\
\cline{1-6}
$2$&$0$&$0$&$7$&$56$\\
\cline{1-5}
$3$&$0$&$0$&$56$\\
\cline{1-4}
$4$&$0$&$0$\\
\cline{1-3}
$5$&$0$\\
\cline{1-2}
\end{tabular}
\begin{tabular}{|c||c|c|c|c|c|c|}
\hline
&$0$&$1$&$2$&$3$&$4$&$5$\\
\hline\hline
$0$&$0$&$0$&$0$&$0$&$0$&$0$\\
\hline
$1$&$0$&$0$&$0$&$0$&$0$\\
\cline{1-6}
$2$&$0$&$0$&$0$&$10$\\
\cline{1-5}
$3$&$0$&$0$&$10$\\
\cline{1-4}
$4$&$0$&$0$\\
\cline{1-3}
$5$&$0$\\
\cline{1-2}
\end{tabular}\\
$n^{{\bm d},{\bm\ell}=(2)}_{g=0}$\hspace{36mm}
$n^{{\bm d},{\bm\ell}=(2)}_{g=1}$\hspace{36mm}
$n^{{\bm d},{\bm\ell}=(2)}_{g=2}$
\caption{The Gopakumar-Vafa invariants $n^{{\bm d},{\bm\ell}}_g$ for $\bm\ell=(2)$.
The asymmetry of the Gopakumar-Vafa invariants in the degrees appear in $n^{(d_+,d_-)=(0,1),{\bm\ell}=(2)}_{g=0}=1$.}
\label{Y2}
\end{center}
\end{table}

The results are given in tables \ref{Y1}, \ref{Y11}, \ref{Y2} for ${\bm\ell}=(1)$, ${\bm\ell}=(1,1)$, ${\bm\ell}=(2)$ respectively up to $d_1+d_2=5$ and table \ref{3boxes} for ${\bm\ell}=(1,1,1)$, ${\bm\ell}=(2,1)$, ${\bm\ell}=(3)$ up to $d_1+d_2=3$.
It is clear that the Gopakumar-Vafa invariants for ${\bm\ell}=(1,1)$ and ${\bm\ell}=(2)$ at $g=0$ are split asymmetrically as $n^{(d_+,d_-)=(0,1),{\bm\ell}=(1,1)}_{g=0}=n^{(d_+,d_-)=(0,1),{\bm\ell}=(2)}_{g=0}=1$ and the asymmetry is even larger for $|{\bm\ell}|=3$ in table \ref{3boxes}.
In determining the BPS indices for $|{\bm\ell}|=2$, we assume the symmetry in exchanging the degrees for higher order of $Q$.
This is because the one-point function in the grand canonical ensemble in appendix \ref{np1pt} is identical for ${\bm\ell}=(1,1)$ and ${\bm\ell}=(2)$ beyond the order of $Q^2$.
Although we only have data with $k=6,8,12$, let us asuume this is the case for any $k$ and $M$.
This implies that $\llangle(\Str U)^2\rrangle^\text{GC}_{k,M}/\llangle 1\rrangle^\text{GC}_{k,M}=(A^{(1,1)}_1+(A^{(1)}_1)^2)/2$ and $\llangle\Str U^2\rrangle^\text{GC}_{k,M}/\llangle 1\rrangle^\text{GC}_{k,M}=(A^{(2)}_1+A^{(1)}_2)/2$ are real and pure imaginary respectively.
Combining with the fact that $\llangle\Str U\rrangle^\text{GC}_{k,M}/\llangle 1\rrangle^\text{GC}_{k,M}=A^{(1)}_1$ is also pure imaginary, we find that $A^{(1,1)}$ and $A^{(2)}$ are themselves real and pure imaginary, which implies the symmetry in exchanging the degrees.

From the geometric viewpoint of local ${\mathbb P}^1\times{\mathbb P}^1$ the two Kahler parameters corresponding to the sizes of ${\mathbb P}^1$ should be symmetric under the exchange.
It is only after we include the Wilson loop insertion indicating one ${\mathbb P}^1$ that the split of the diagonal BPS indices happens.

\begin{table}[ht!]
\begin{center}
\begin{tabular}{|c||c|c|c|c|}
\hline
&$0$&$1$&$2$&$3$\\
\hline\hline
$0$&$0$&$0$&$0$&$0$\\
\hline
$1$&$1$&$2$&$3$\\
\cline{1-4}
$2$&$1$&$6$\\
\cline{1-3}
$3$&$0$\\
\cline{1-2}
\end{tabular}
\begin{tabular}{|c||c|c|c|c|}
\hline
&$0$&$1$&$2$&$3$\\
\hline\hline
$0$&$0$&$0$&$0$&$0$\\
\hline
$1$&$1$&$2$&$3$\\
\cline{1-4}
$2$&$1$&$6$\\
\cline{1-3}
$3$&$0$\\
\cline{1-2}
\end{tabular}
\begin{tabular}{|c||c|c|c|c|}
\hline
&$0$&$1$&$2$&$3$\\
\hline\hline
$0$&$0$&$0$&$0$&$0$\\
\hline
$1$&$1$&$2$&$3$\\
\cline{1-4}
$2$&$1$&$6$\\
\cline{1-3}
$3$&$0$\\
\cline{1-2}
\end{tabular}\\
$n^{{\bm d},{\bm\ell}=(1,1,1)}_{g=0}$\hspace{18mm}
$n^{{\bm d},{\bm\ell}=(2,1)}_{g=0}$\hspace{18mm}
$n^{{\bm d},{\bm\ell}=(3)}_{g=0}$
\caption{The Gopakumar-Vafa invariants $n^{{\bm d},{\bm\ell}}_g$ for $\bm\ell=(3)$, $\bm\ell=(2,1)$ and $\bm\ell=(1,1,1)$ respectively.}
\label{3boxes}
\end{center}
\end{table}

\section{Conclusion and discussions}\label{conclusion}

In this paper, we have introduced the two-point function and studied it numerically.
Though we have defined the two-point function so that it does not decompose to the one-point functions trivially, after our full analysis, we have found two unexpected relations to the one-point functions.
One of them relates the non-trivial two-point function to the trivial one which further reduces to a combination of the one-point functions through the Littlewood-Richardson rule.
The other relates the imaginary part of the two-point function to the one-point functions in the representation with the box number smaller than the total number by two.
We have also revisited the one-point function and identified how the diagonal BPS indices split by the degree difference asymmetrically.

Apparently there are many questions related to our result.

The first most important one would be the physical origin of the two-point function.
In this paper we have defined the two-point function with two characters of the opposite charges in the matrix model.
It is of course desirable to understand how the two-point function arises from the correlation function in the ABJM theory.
We would like to identify it as the two-point function of the physical Wilson loops in the ABJM theory and derive our two-point function from the localization techniques.

Although we have the general expression for the conjugate relation in section \ref{superposerel}, the general expression for the descent relation in section \ref{interfererel} is missing, where we have only proposed some relations in the main text and in appendix \ref{interference}.
Also, the physical interpretation of the two relations we have found is unclear.
We can imagine that the two relations reflect the topological nature and the orientifold or symplectic nature of the ABJM matrix model.
Especially in appendix \ref{heisenberg}, we make a proposal on the relation between the two-point function with the main phase removed $\im e^{-i\Theta^{Y,Z}_{k,M}}\langle s_Y\bar s_Z\rangle^\text{GC}_{k,M}$ and a bracket between $s_Y$ and $s_Z$.
We hope to elaborate the interpretation.

It is surprising to us that the two-point function turns out to closely relate to the representation theory of the supergroup U$(N_1|N_2)$ in the Fermi gas formalism.
It is, however, still unclear, what role the composite Young diagram appearing in the Fermi gas formalism plays in the representation theory.

Our definition of the non-trivial two-point functions in section \ref{definition} has a direct generalization to other ${\cal N}=4$ superconformal Chern-Simons matrix models of type $\widehat A$ \cite{HM,MN1,MN2,MN3,HHO,MNN,MNY}.
Namely, for the circular quiver gauge group $\prod_{i}$U$(N_i)$ with an even number of nodes, the Fermi gas formalism works naturally when the arguments of the inserted supersymmetric Schur functions appear reversely for the adjacent nodes as in $s_{Y_i}(x^{(N_i)}|x^{(N_{i+1})})$ and $s_{Y_{i+1}}((x^{-1})^{(N_{i+1})}|(x^{-1})^{(N_{i+2})})$.
In terms of the original gauge theory, we expect that, for the multiple insertion of the Wilson loops in the ${\cal N}=4$ superconformal Chern-Simons theories, the correlation functions preserve half of the supersymmetries only when the charges of the two adjacent loop insertions are reverse.
We hope to study this fact following the discussions in \cite{OWZ,CDT,GLMPS,BGLMPS,LMPZ}.

\appendix

\section{Determinantal formulas}\label{determinant}

In this appendix, we summarize several determinantal formulas which are useful in our derivation of the Fermi gas formalism in section \ref{derivation}.
The first one is collected from \cite{MNN}, generalizing the previous simpler version with $M_2=0$ in \cite{MM}.
The remaining two are collected from \cite{MM}.

\noindent
{\it Formula 1.}
Let $(\phi_i)_{1\le i\le N+R_1}$ and $(\psi_j)_{1\le j\le N+R_2}$ be arrays of functions of $x$ and let $(\xi_{ik})_{\begin{subarray}{c}1\le i\le N+R_1\\N+1\le k\le N+R_1\end{subarray}}$ and $(\eta_{lj})_{\begin{subarray}{c}N+1\le l\le N+R_2\\1\le j\le N+R_2\end{subarray}}$ be arrays of constants. Then, we have
\begin{align*}
&\int\frac{d^Nx}{N!}
\det\begin{pmatrix}\bigl(\phi_i(x_k)\bigr)
_{\begin{subarray}{c}1\le i\le N+R_1\\1\le k\le N\end{subarray}}&
\bigl(\xi_{ik}\bigr)
_{\begin{subarray}{c}1\le i\le N+R_1\\N+1\le k\le N+R_1\end{subarray}}
\end{pmatrix}
\det\begin{pmatrix}(\psi_j(x_l))
_{\begin{subarray}{c}1\le l\le N\\1\le j\le N+R_2\end{subarray}}\\
\bigl(\eta_{lj}\bigr)
_{\begin{subarray}{c}N+1\le l\le N+R_2\\1\le j\le N+R_2\end{subarray}}
\end{pmatrix}\\
&=(-1)^{R_1R_2}\det\begin{pmatrix}\bigl(\phi_i\circ\psi_j\bigr)
_{\begin{subarray}{c}1\le i\le N+R_1\\1\le j\le N+R_2\end{subarray}}&
\bigl(\xi_{ik}\bigr)
_{\begin{subarray}{c}1\le i\le N+R_1\\N+1\le k\le N+R_1\end{subarray}}\\
\bigl(\eta_{lj}\bigr)
_{\begin{subarray}{c}N+1\le l\le N+R_2\\1\le j\le N+R_2\end{subarray}}&
(0)
_{\begin{subarray}{c}N+1\le l\le N+R_2\\N+1\le k\le N+R_1\end{subarray}}
\end{pmatrix}
\end{align*}
with
\begin{align*}
\phi_i\circ\psi_j=\int dx\phi_i(x)\psi_j(x).
\end{align*}

\bigskip

\noindent
{\it Formula 2.}
Let $A(x,x')$ be a function of $x$ and $x'$, $(B_k(x))_{\begin{subarray}{c}1\le k\le M\end{subarray}}$ and $(C_l(x))_{\begin{subarray}{c}1\le l\le M\end{subarray}}$ be arrays of functions of $x$ and $(D_{k,l})_{\begin{subarray}{c}1\le k\le M\\1\le l\le M\end{subarray}}$ be an array of constants.
Then, we have
\begin{align}
\sum_{N=0}^\infty\int\frac{d^Nx}{N!}
\det\begin{pmatrix}A(x_i,x_j)&B_k(x_i)\\C_l(x_j)&D_{k,l}\end{pmatrix}
=\Det\begin{pmatrix}1+A&B\\C&D\end{pmatrix},
\end{align}
where $\det$ on the left-hand side is an ordinary determinant in dimension $N+M$, while $\Det$ on the right-hand side is a determinant combined with the Fredholm determinant in infinite dimension of the function space and an ordinary determinant in dimension $M$.

\bigskip

\noindent
{\it Formula 3.}
For a square matrix containing two diagonal blocks of square matrices $A$ and $D$, the determinant formula holds,
\begin{align}
\det\begin{pmatrix}A&B\\C&D\end{pmatrix}
=\det A\det(D-CA^{-1}B).
\end{align}

\section{Lowest component}\label{lowest}

After establishing the Fermi gas formalism for the two-point function, let us study the lowest component of it.
For this purpose, we fix $M\ge 0$ and $R'\le R$ and study the lowest components of each ingredient first,
\begin{align}
\Xi_k(w)&=1+{\cal O}(w),\nonumber\\
K_k^{(a'|l)}(w)&=\int\frac{d\nu}{2\pi}e^{-\frac{ik}{4\pi}\nu^2}e^{(a'+l)\nu}+{\cal O}(w)
=\frac{e^{-\frac{\pi i}{k}(a'+l)^2}}{\sqrt{ik}}+{\cal O}(w),\nonumber\\
H_k^{(a|l)}(w)&=w\int\frac{d\nu}{2\pi}\frac{d\mu}{2\pi}e^{l\nu}e^{-\frac{ik}{4\pi}\nu^2}
\frac{1}{2\cosh\frac{\nu-\mu}{2}}e^{\frac{ik}{4\pi}\mu^2}e^{a\mu}+{\cal O}(w^2)
=\frac{e^{\frac{\pi i}{k}(a^2-l^2)}}{2k\cos\frac{\pi(a+l)}{k}}w+{\cal O}(w^2).
\end{align}
Then, the two-point function becomes
\begin{align}
&\langle s_Y\bar s_{Y'}\rangle^\text{GC}_{k,M}(z)
=i^{\frac{1}{2}M^2}e^{\frac{\pi i}{k}(\sum a^2-\sum l^2-\sum a'^2+\sum l'^2)}
\nonumber\\
&\quad\times\det\begin{pmatrix}
\displaystyle\biggl[\frac{1}{2k\cos\frac{\pi(a'+l')}{k}}w
+{\cal O}(w^2)\biggr]_{(R'+M)\times R'}&
\displaystyle\biggl[\frac{e^{-\frac{2\pi i}{k}la'}}{\sqrt{ik}}
+{\cal O}(w)\biggr]_{(R'+M)\times(M+R)}\\
\displaystyle\biggl[-\frac{e^{\frac{2\pi i}{k}al'}}{\sqrt{-ik}}w
+{\cal O}(w^2)\biggr]_{R\times R'}&
\displaystyle\biggl[\frac{1}{2k\cos\frac{\pi(a+l)}{k}}w
+{\cal O}(w^2)\biggr]_{R\times(M+R)}
\end{pmatrix}.
\end{align}
Hereafter let us drop the symbol denoting higher components.
Suppose we bring the factor $w$ in the lower blocks out of the determinant.
If the determinant is non-vanishing after setting all components depending on $w$ to be zero, then the lowest component is
\begin{align}
&\langle s_Y\bar s_{Y'}\rangle^\text{GC}_{k,M}(z)
=i^{\frac{1}{2}M^2}e^{\frac{\pi i}{k}(\sum a^2-\sum l^2-\sum a'^2+\sum l'^2)}
\frac{w^R(-1)^{R'}}{\sqrt{ik}^{M+R}\sqrt{-ik}^R}\nonumber\\
&\quad\times(-1)^{(R'+M)R+R'(M+R)+\frac{1}{2}R'(R'-1)+\frac{1}{2}(M+R')(M+R'-1)}\Delta,
\end{align}
with $\Delta$ being an abbreviation for the determinant part
\begin{align}
\Delta=
\det\begin{pmatrix}\displaystyle\biggl[\frac{1}{2\cos\frac{\pi(a+l)}{k}}\biggr]_{R\times(M+R)}&
\displaystyle\Bigl[e^{\frac{2\pi i}{k}al'}\Bigr]_{R\times R'}\\
\displaystyle\Bigl[e^{-\frac{2\pi i}{k}la'}\Bigr]_{(R'+M)\times(M+R)}&
\displaystyle[0]_{(R'+M)\times R'}\end{pmatrix}.
\end{align}
Note that we have exchanged the two row blocks and the two column blocks and at the same time rearranged the arm lengths $\{-l'\}$ and the leg lengths $\{-a'\}$ of the Young diagram $Y'$ to be in the standard order of a single Young diagram.
In fact the determinant is non-vanishing in general for $R'\le R$ and computable, which is clear if we rewrite the determinant as
\begin{align}
\Delta=e^{\frac{\pi i}{k}(-\sum a+\sum l)}
\det\begin{pmatrix}
\displaystyle\biggl[\frac{1}{e^{-\frac{2\pi i}{k}a}+e^{\frac{2\pi i}{k}l}}\biggr]&
\displaystyle\Bigl[e^{-\frac{2\pi i}{k}a(-l'-\frac{1}{2})}\Bigr]\\
\displaystyle\Bigl[e^{\frac{2\pi i}{k}l(-a'-\frac{1}{2})}\Bigr]&[0]
\end{pmatrix},
\end{align}
and apply the transpose of \eqref{Ybar} along with \eqref{Zbar} with $x_i^{-1}=e^{-\frac{2\pi i}{k}a_i}$ and $y_j^{-1}=e^{\frac{2\pi i}{k}l_j}$.
Then, we find
\begin{align}
\Delta
&=e^{\frac{\pi i}{k}(-\sum a+\sum l)}(-1)^{R'+MR}
\frac{\prod_{i<i'}^R(e^{-\frac{2\pi i}{k}a_{i}}-e^{-\frac{2\pi i}{k}a_{i'}})
\prod_{j<j'}^{M+R}(e^{\frac{2\pi i}{k}l_{j}}-e^{\frac{2\pi i}{k}l_{j'}})}
{\prod_{i=1}^R\prod_{j=1}^{M+R}(e^{-\frac{2\pi i}{k}a_i}+e^{\frac{2\pi i}{k}l_j})}
s_{Y'}(e^{-\frac{2\pi i}{k}a}|e^{\frac{2\pi i}{k}l})\nonumber\\
&=i^{-\frac{1}{2}R(R-1)+\frac{1}{2}(M+R)(M+R-1)}(-1)^{R'+MR}
e^{\frac{\pi i}{k}M(\sum a+\sum l)}
\nonumber\\
&\quad\times
\frac{\prod_{i<i'}^R2\sin\frac{\pi(a_i-a_{i'})}{k}
\prod_{j<j'}^{M+R}2\sin\frac{\pi(l_j-l_{j'})}{k}}
{\prod_{i=1}^R\prod_{j=1}^{M+R}2\cos\frac{\pi(a_i+l_j)}{k}}
s_{Y'}(e^{-\frac{2\pi i}{k}a}|e^{\frac{2\pi i}{k}l}).
\end{align}
To summarize for now, the two-point function is given by
\begin{align}
\langle s_Y\bar s_{Y'}\rangle^\text{GC}_{k,M}(z)
&=e^{\frac{\pi i}{k}(\sum a^2-\sum l^2-\sum a'^2+\sum l'^2)}
e^{\frac{\pi i}{k}M(\sum a+\sum l)}
\nonumber\\
&\quad\times\frac{z^R}{k^{R+\frac{M}{2}}}
\frac{\prod_{i<i'}^R2\sin\frac{\pi(a_i-a_{i'})}{k}
\prod_{j<j'}^{M+R}2\sin\frac{\pi(l_j-l_{j'})}{k}}
{\prod_{i=1}^R\prod_{j=1}^{M+R}2\cos\frac{\pi(a_i+l_j)}{k}}
s_{Y'}(e^{-\frac{2\pi i}{k}a}|e^{\frac{2\pi i}{k}l}).
\label{posMR}
\end{align}

Let us take a detour to study the phase factor.
For this purpose we compute in two ways the sum of the shifted contents $j-i-M$ over all of the boxes, where the arrows representing the arm lengths or the leg lengths pierce (see figure \ref{young}).
On one hand we sum up the shifted contents along the arrows defining the arm lengths and the leg lengths, while on the other hand we compute the same quantity separately for the Young diagram part (the yellow/cyan part for $Y$/$Y'$ in figure \ref{young}) and the triangular part where only the arrows defining the auxiliary arm or leg lengths pierce (the green part in figure \ref{young}),
\begin{align}
\sum_{i=1}^R\sum_{j=1}^{a_i-\frac{1}{2}}j-\sum_{j=1}^{M+R}\sum_{i=1}^{l_j-\frac{1}{2}}i
=\biggl(\sum_{Y}+\sum_{\triangle}\biggr)(j-i-M),
\end{align}
where $\triangle$ denote the sum over the green triangular region.
Then we find that
\begin{align}
\frac{1}{2}\biggl(\sum_{i=1}^Ra_i^2-\sum_{j=1}^{M+R}l_j^2+\frac{M}{4}\biggr)
=c^Y-M|Y|-\frac{1}{6}(M^3-M),
\end{align}
for $Y$, along with
\begin{align}
\frac{1}{2}\biggl(\sum_{j=1}^{R'}l_j'^2-\sum_{i=1}^{M+R'}a_i'^2+\frac{M}{4}\biggr)
=c^{Y'}-M|Y'|-\frac{1}{6}(M^3-M),
\end{align}
for $Y'$, which implies that the first line in \eqref{posMR} is
\begin{align}
&\sum a^2-\sum l^2-\sum a'^2+\sum l'^2+M(\sum a+\sum l)\nonumber\\
&\quad=-\frac{1}{6}(M^3-M)+2c^Y-M|Y|+2c^{Y'}-2M|Y'|,
\end{align}
resembling a lot with the main phase factor $\Theta^{Y,Y'}_{k,M}$ \eqref{twophase}.

Finally the two-point function with the main phase factor removed can be expressed as
\begin{align}
e^{-i\Theta^{Y,Y'}_{k,M}}\langle s_Y\bar s_{Y'}\rangle^\text{GC}_{k,M}(z)
=\frac{z^R}{k^{R+\frac{M}{2}}}
\frac{\prod_{i<i'}^{R}2\sin\frac{\pi(a_i-a_{i'})}{k}
\prod_{j<j'}^{M+R}2\sin\frac{\pi(l_j-l_{j'})}{k}}
{\prod_{i=1}^{R}\prod_{j=1}^{M+R}2\cos\frac{\pi(a_i+l_j)}{k}}
e^{-\frac{\pi i}{k}M|Y'|}
s_{Y'}(e^{-\frac{2\pi i}{k}a}|e^{\frac{2\pi i}{k}l}).
\label{lowestR}
\end{align}
On the other hand, when $R'\ge R$ the expression is
\begin{align}
e^{-i\Theta^{Y,Y'}_{k,M}}\langle s_Y\bar s_{Y'}\rangle^\text{GC}_{k,M}(z)
=\frac{z^{R'}}{k^{R'+\frac{M}{2}}}
\frac{\prod_{j<j'}^{R'+M}2\sin\frac{\pi(l'_j-l'_{j'})}{k}
\prod_{i<i'}^{R'}2\sin\frac{\pi(a'_i-a'_{i'})}{k}}
{\prod_{j=1}^{R'+M}\prod_{i=1}^{R'}2\cos\frac{\pi(a'_i+l'_j)}{k}}
e^{-\frac{\pi i}{k}M|Y|}
s_{Y}(e^{\frac{2\pi i}{k}l'}|e^{-\frac{2\pi i}{k}a'}),
\label{lowestR'}
\end{align}
which is obtained by exchanging $Y$ and $Y'$.
The expressions of \eqref{lowestR} and \eqref{lowestR'} are valid also for $M\le 0$ with the understanding $M=-\bar M$, $R=\bar M+\bar R$, $M+R=\bar R$, $R'=\bar R'+\bar M$, $R'+M=\bar R'$.

For the case of one-point functions with $Y'=\varnothing$, the expression reduces to
\begin{align}
e^{-i\Theta^Y_{k,M}}\langle s_Y\rangle^\text{GC}_{k,M}(z)
=\frac{z^R}{k^{R+\frac{M}{2}}}
\frac{\prod_{i<i'}^{R}2\sin\frac{\pi(a_i-a_{i'})}{k}
\prod_{j<j'}^{M+R}2\sin\frac{\pi(l_j-l_{j'})}{k}}
{\prod_{i=1}^{R}\prod_{j=1}^{M+R}2\cos\frac{\pi(a_i+l_j)}{k}}.
\end{align}

\section{Conjugate relation}\label{superpose}

In this appendix we shall summarize some numerical data to convince the reader of our proposal of the conjugate relation discussed in section \ref{superposerel}.
Due to the vast data, we introduce the abbreviation
\begin{align}
W^{Y,Z}_{k,M}(N)=e^{-i\Theta^{Y,Z}_{k,M}}\langle s_Y\bar s_Z\rangle_k(N,N+M),\quad
W^{Y}_{k,M}(N)=e^{-i\Theta^{Y}_{k,M}}\langle s_Y\rangle_k(N,N+M),
\end{align}
and only record those with $k=6$ and $M=0,1,2$ in this appendix.

On one hand, the exact values of the two-point function $W^{\yng(1),\yng(1)}_{k,M}(N)$ for $k=6$ are given by
\begin{align}
&W^{\yng(1),\yng(1)}_{6,0}(1)=\frac{1}{6},\quad
W^{\yng(1),\yng(1)}_{6,0}(2)=\frac{1}{432},\quad
W^{\yng(1),\yng(1)}_{6,0}(3)=\frac{378+96\sqrt{3}\pi-91\pi^2}{93312\pi^2},
\nonumber\\
&W^{\yng(1),\yng(1)}_{6,1}(0)
=\frac{1+i\sqrt{3}}{2\sqrt{6}},\quad
W^{\yng(1),\yng(1)}_{6,1}(1)
=\frac{9i+3\sqrt{3}+(11-i\sqrt{3})\pi}{216\sqrt{2}\pi},\nonumber\\
&\quad
W^{\yng(1),\yng(1)}_{6,1}(2)
=\frac{432(-3i-\sqrt{3})+72(43-17i\sqrt{3})\pi+(807i-499\sqrt{3})\pi^2}{186624\sqrt{2}\pi^2},
\nonumber\\
&\quad
W^{\yng(1),\yng(1)}_{6,1}(3)
=\bigl(3240(-3i-\sqrt{3})+2592(-19+5i\sqrt{3})\pi+(4347i-5463\sqrt{3})\pi^2\nonumber\\
&\qquad
+(8185-1931i\sqrt{3})\pi^3\bigr)\big/\bigl(20155392\sqrt{2}\pi^3\bigr),\nonumber\\
&W^{\yng(1),\yng(1)}_{6,2}(0)
=\frac{3i+\sqrt{3}}{12},\quad
W^{\yng(1),\yng(1)}_{6,2}(1)
=\frac{-36+12i\sqrt{3}+(5i+7\sqrt{3})\pi}{864\pi},\nonumber\\
&\quad
W^{\yng(1),\yng(1)}_{6,2}(2)
=\frac{162i+54\sqrt{3}+(324-108i\sqrt{3})\pi+(45i-65\sqrt{3})\pi^2}{62208\pi^2},\nonumber\\
&\quad
W^{\yng(1),\yng(1)}_{6,2}(3)
=\bigl(-12636+4212i\sqrt{3}+(2835i+3969\sqrt{3})\pi+(24030-1098i\sqrt{3})\pi^2\nonumber\\
&\qquad
+(83i-4583\sqrt{3})\pi^3\bigr)\big/\bigl(20155392\pi^3\bigr).
\end{align}
On the other hand, the exact values of the one-point function $W^{\yng(2)}_{k,M}(N)$ for $k=6$ are given by
\begin{align}
&W^{\yng(2)}_{6,0}(1)=\frac{1}{6},\quad
W^{\yng(2)}_{6,0}(2)=\frac{1}{432},\quad
W^{\yng(2)}_{6,0}(3)=\frac{378+96\sqrt{3}\pi-91\pi^2}{93312\pi^2},\nonumber\\
&W^{\yng(2)}_{6,1}(1)=\frac{1}{18\sqrt{2}},\quad
W^{\yng(2)}_{6,1}(2)=\frac{45\sqrt{2}-8\sqrt{6}\pi}{3888\pi},\quad
W^{\yng(2)}_{6,1}(3)=\frac{-1728-192\sqrt{3}\pi+281\pi^2}{559872\sqrt{2}\pi^2},
\nonumber\\
&W^{\yng(2)}_{6,2}(0)=\frac{1}{6},\quad
W^{\yng(2)}_{6,2}(1)=\frac{12\sqrt{3}-\pi}{432\pi} ,\quad
W^{\yng(2)}_{6,2}(2)=\frac{54-108\sqrt{3}\pi+55\pi^2}{31104\pi^2},\nonumber\\
&\quad
W^{\yng(2)}_{6,2}(3)=\frac{4212\sqrt{3}-567\pi-4554\sqrt{3}\pi^2+2333\pi^3}{10077696\pi^3},
\end{align}
while the exact values of the one-point function $W^{\yng(1,1)}_{k,M}(N)$ for $k=6$ are given by
\begin{align}
&W^{\yng(1,1)}_{6,0}(1)=\frac{1}{6},\quad
W^{\yng(1,1)}_{6,0}(2)=\frac{1}{432},\quad
W^{\yng(1,1)}_{6,0}(3)=\frac{378+96\sqrt{3}\pi-91\pi^2}{93312\pi^2},\nonumber\\
&W^{\yng(1,1)}_{6,1}(0)=\frac{1}{\sqrt{6}},\quad
W^{\yng(1,1)}_{6,1}(1)=\frac{3\sqrt{6}+5\sqrt{2}\pi}{216\pi},\quad
W^{\yng(1,1)}_{6,1}(2)=\frac{-432+312\sqrt{3}\pi-115\pi^2}{31104\sqrt{6}\pi^2},\nonumber\\
&\quad
W^{\yng(1,1)}_{6,1}(3)=\frac{-9720-18144\sqrt{3}\pi-6021\pi^2+3127\sqrt{3}\pi^3}
{10077696\sqrt{6}\pi^3}.
\end{align}
These exact values satisfy the conjugate relation \eqref{superposeid}.

\section{Descent relation}\label{interfere}

\subsection{Explicit relations}\label{interference}

In this appendix we list up the explicit forms of the descent relation discussed in section \ref{interfererel}.
We adopt the abbreviated notation we have introduced in \eqref{generalinterfere} with the tacit understanding of mod $e^{i\Theta^{Y,Z}}{\mathbb R}$.
For the case of the two-point functions with two, three and four boxes in total, the relations all fall into the patterns \eqref{generalbox}, \eqref{hooksym}, \eqref{antisymsym} discussed in the main text,
\begin{align}
&\langle s_{\yng(1)}\bar s_{\yng(1)}\rangle
\sim\langle 1\rangle,\quad
\langle s_{\yng(2)}\bar s_{\yng(1)}\rangle
\sim\langle s_{\yng(1)}\rangle,\quad
\langle s_{\yng(1,1)}\bar s_{\yng(1)}\rangle
\sim\langle s_{\yng(1)}\rangle,
\nonumber\\
&\langle s_{\yng(3)}\bar s_{\yng(1)}\rangle
\sim\langle s_{\yng(2)}\rangle,\quad
\langle s_{\yng(2,1)}\bar s_{\yng(1)}\rangle
\sim\langle s_{\yng(2)}\rangle+\langle s_{\yng(1,1)}\rangle,\quad
\langle s_{\yng(1,1,1)}\bar s_{\yng(1)}\rangle
\sim\langle s_{\yng(1,1)}\rangle,
\nonumber\\
&\langle s_{\yng(2)}\bar s_{\yng(2)}\rangle
\sim q\langle s_{\yng(2)}\rangle+q^{-1}\langle s_{\yng(1,1)}\rangle,\quad
\langle s_{\yng(1,1)}\bar s_{\yng(2)}\rangle
\sim\langle s_{\yng(2)}\rangle+\langle s_{\yng(1,1)}\rangle,\quad
\langle s_{\yng(1,1)}\bar{s}_{\yng(1,1)}\rangle
\sim q\langle s_{\yng(2)}\rangle+q^{-1}\langle s_{\yng(1,1)}\rangle.
\end{align}
For the case of the two-point functions with five boxes in total, the expressions are not unique due to the relation \eqref{onepointthree} among the one-point functions with three boxes.
We choose a representative set of the coefficients for our expressions.
\begin{align}
&\langle s_{\yng(4)}\bar s_{\yng(1)}\rangle
\sim\langle s_{\yng(3)}\rangle,\quad
\langle s_{\yng(3,1)}\bar s_{\yng(1)}\rangle
\sim\langle s_{\yng(3)}\rangle+\langle s_{\yng(2,1)}\rangle,\quad
\langle s_{\yng(2,2)}\bar s_{\yng(1)}\rangle
\sim\langle s_{\yng(2,1)}\rangle,
\nonumber\\
&\langle s_{\yng(2,1,1)}\bar s_{\yng(1)}\rangle
\sim\langle s_{\yng(2,1)}\rangle+\langle s_{\yng(1,1,1)}\rangle,\quad
\langle s_{\yng(1,1,1,1)}\bar s_{\yng(1)}\rangle
\sim\langle s_{\yng(1,1,1)}\rangle,
\nonumber\\
&\langle s_{\yng(3)}\bar s_{\yng(2)}\rangle
\sim q^2\langle s_{\yng(3)}\rangle+q^{-1}\langle s_{\yng(2,1)}\rangle,\quad
\langle s_{\yng(2,1)}\bar s_{\yng(2)}\rangle
\sim(q+1+q^{-1})\langle s_{\yng(2,1)}\rangle,\quad
\langle s_{\yng(1,1,1)}\bar s_{\yng(2)}\rangle
\sim\langle s_{\yng(2,1)}\rangle+\langle s_{\yng(1,1,1)}\rangle,
\nonumber\\
&\langle s_{\yng(3)}\bar{s}_{\yng(1,1)}\rangle
\sim\langle s_{\yng(3)}\rangle
+\langle s_{\yng(2,1)}\rangle,\quad
\langle s_{\yng(2,1)}\bar{s}_{\yng(1,1)}\rangle
\sim(q+1+q^{-1})\langle s_{\yng(2,1)}\rangle,\quad
\langle s_{\yng(1,1,1)}\bar{s}_{\yng(1,1)}\rangle
\sim q\langle s_{\yng(2,1)}\rangle+q^{-2}\langle s_{\yng(1,1,1)}\rangle.
\label{5descent}
\end{align}
Note that the third relation can be alternatively expressed as $\langle s_{\yng(2,2)}\bar s_{\yng(1)}\rangle\to q\langle s_{\yng(3)}\rangle+q^{-1}\langle s_{\yng(1,1,1)}\rangle$ for example due to the relation \eqref{onepointthree} among the one-point functions.
Also note that the relations for $\langle s_{\yng(2,1)}\bar s_{\yng(2)}\rangle$ and $\langle s_{\yng(2,1)}\bar{s}_{\yng(1,1)}\rangle$ are not included in the patterns \eqref{generalbox}, \eqref{hooksym}, \eqref{antisymsym} and are new.
For the case of the two-point functions with five boxes in total, again the expressions are not unique and we choose a representative set of the coefficients.
The expressions with $(|Y|,|Z|)=(4,2)$ are
\begin{align}
&\langle s_{\yng(4)}\bar{s}_{\yng(2)}\rangle
\sim q^3\langle s_{\yng(4)}\rangle+q^{-1}\langle s_{\yng(3,1)}\rangle,\nonumber\\
&\langle s_{\yng(3,1)}\bar{s}_{\yng(2)}\rangle
\sim(q^2+q+1)\langle s_{\yng(4)}\rangle
+q^2\langle s_{\yng(3,1)}\rangle
+(q^{-1}+q^{-2})\langle s_{\yng(2,1,1)}\rangle,\nonumber\\
&\langle s_{\yng(2,2)}\bar{s}_{\yng(2)}\rangle
\sim(q^3+q)\langle s_{\yng(4)}\rangle
+(1-q^{-1})\langle s_{\yng(2,2)}\rangle
+(1+q^{-2})\langle s_{\yng(2,1,1)}\rangle,\nonumber\\
&\langle s_{\yng(2,1,1)}\bar{s}_{\yng(2)}\rangle
\sim(q+1)\langle s_{\yng(3,1)}\rangle
+q\langle s_{\yng(2,1,1)}\rangle
+(q^{-1}+q^{-2}+q^{-3})\langle s_{\yng(1,1,1,1)}\rangle,\quad
\langle s_{\yng(1,1,1,1)}\bar{s}_{\yng(2)}\rangle
\sim\langle s_{\yng(2,1,1)}\rangle
+\langle s_{\yng(1,1,1,1)}\rangle,\nonumber\\
&\langle s_{\yng(4)}\bar{s}_{\yng(1,1)}\rangle
\sim\langle s_{\yng(4)}\rangle
+\langle s_{\yng(3,1)}\rangle,\quad
\langle s_{\yng(3,1)}\bar{s}_{\yng(1,1)}\rangle
\sim(q^3+q^2+q)\langle s_{\yng(4)}\rangle
+q^{-1}\langle s_{\yng(3,1)}\rangle
+(1+q^{-1})\langle s_{\yng(2,1,1)}\rangle,\nonumber\\
&\langle s_{\yng(2,2)}\bar{s}_{\yng(1,1)}\rangle
\sim(q^2+1)\langle s_{\yng(3,1)}\rangle
+(-q+1)\langle s_{\yng(2,2)}\rangle
+(q^{-1}+q^{-3})\langle s_{\yng(1,1,1,1)}\rangle,\nonumber\\
&\langle s_{\yng(2,1,1)}\bar{s}_{\yng(1,1)}\rangle
\sim(q^2+q)\langle s_{\yng(3,1)}\rangle
+q^{-2}\langle s_{\yng(2,1,1)}\rangle
+(1+q^{-1}+q^{-2})\langle s_{\yng(1,1,1,1)}\rangle,\nonumber\\
&\langle s_{\yng(1,1,1,1)}\bar{s}_{\yng(1,1)}\rangle
\sim q\langle s_{\yng(2,1,1)}\rangle
+q^{-3}\langle s_{\yng(1,1,1,1)}\rangle,
\label{descentYZ42}
\end{align}
while the expressions with $(|Y|,|Z|)=(3,3)$ are
\begin{align}
&\langle s_{\yng(3)}\bar{s}_{\yng(3)}\rangle
\sim q^4\langle s_{\yng(4)}\rangle
+\langle s_{\yng(3,1)}\rangle
+q^{-2}\langle s_{\yng(2,2)}\rangle,\nonumber\\
&\langle s_{\yng(2,1)}\bar{s}_{\yng(3)}\rangle
\sim(q^2+q+1)\langle s_{\yng(4)}\rangle
+q^2\langle s_{\yng(3,1)}\rangle
+(q^{-1}+q^{-2})\langle s_{\yng(2,1,1)}\rangle,\nonumber\\
&\langle s_{\yng(1,1,1)}\bar{s}_{\yng(3)}\rangle
\sim\langle s_{\yng(3,1)}\rangle
+\langle s_{\yng(2,1,1)}\rangle,\nonumber\\
&\langle s_{\yng(2,1)}\bar{s}_{\yng(2,1)}\rangle
\sim(q^2+2)\langle s_{\yng(3,1)}\rangle
+(q^2+1+q^{-2})\langle s_{\yng(2,2)}\rangle
+(2+q^{-2})\langle s_{\yng(2,1,1)}\rangle,\nonumber\\
&\langle s_{\yng(1,1,1)}\bar{s}_{\yng(2,1)}\rangle
\sim(q^2+q)\langle s_{\yng(3,1)}\rangle
+q^{-2}\langle s_{\yng(2,1,1)}\rangle
+(1+q^{-1}+q^{-2})\langle s_{\yng(1,1,1,1)}\rangle,\nonumber\\
&\langle s_{\yng(1,1,1)}\bar{s}_{\yng(1,1,1)}\rangle
\sim q^2\langle s_{\yng(2,2)}\rangle
+\langle s_{\yng(2,1,1)}\rangle
+q^{-4}\langle s_{\yng(1,1,1,1)}\rangle.
\label{descentYZ33}
\end{align}

\subsection{Ambiguities}\label{ambiguity}

In this appendix we list the relations among the one-point functions with the same box number, which cause the ambiguities when we express the imaginary part of the two-point function in terms of the one-point functions.
We adopt the abbreviated notation \eqref{ambiX} and drop mod $e^{i\Theta^{(n)}}{\mathbb R}$ for simplicity.
For the case with less than five boxes we find
\begin{align}
&q\langle s_{\yng(3)}\rangle-\langle s_{\yng(2,1)}\rangle+q^{-1}\langle s_{\yng(1,1,1)}\rangle
\sim 0,\nonumber\\
&(q^3+q^2)\langle s_{\yng(4)}\rangle
-q\langle s_{\yng(3,1)}\rangle
-q\langle s_{\yng(2,2)}\rangle
+\langle s_{\yng(2,1,1)}\rangle
\sim 0,\nonumber\\
&\langle s_{\yng(3,1)}\rangle
-q^{-1}\langle s_{\yng(2,2)}\rangle
-q^{-1}\langle s_{\yng(2,1,1)}\rangle
+(q^{-2}+q^{-3})\langle s_{\yng(1,1,1,1)}\rangle
\sim 0,
\end{align}
while for the case with five boxes we find
\begin{align}
&(q^5+q^4)\langle s_{\yng(5)}\rangle
-(q^3+q^2)\langle s_{\yng(4,1)}\rangle
+q\langle s_{\yng(3,2)}\rangle
+q\langle s_{\yng(3,1,1)}\rangle
-\langle s_{\yng(2,2,1)}\rangle
\sim 0,\nonumber\\
&q^3\langle s_{\yng(4,1)}\rangle
-(q^2+q)\langle s_{\yng(3,2)}\rangle
+(1+q^{-1})\langle s_{\yng(2,2,1)}\rangle
-q^{-2}\langle s_{\yng(2,1,1,1)}\rangle
\sim 0,\nonumber\\
&(q^2+q)\langle s_{\yng(4,1)}\rangle
-\langle s_{\yng(3,2)}\rangle
-\langle s_{\yng(3,1,1)}\rangle
-\langle s_{\yng(2,2,1)}\rangle
+(q^{-1}+q^{-2})\langle s_{\yng(2,1,1,1)}\rangle
\sim 0,\nonumber\\
&q^2\langle s_{\yng(4,1)}\rangle
-(q+1)\langle s_{\yng(3,2)}\rangle
+(q^{-1}+q^{-2})\langle s_{\yng(2,2,1)}\rangle
-q^{-3}\langle s_{\yng(2,1,1,1)}\rangle
\sim 0,\nonumber\\
&\langle s_{\yng(3,2)}\rangle
-q^{-1}\langle s_{\yng(3,1,1)}\rangle
-q^{-1}\langle s_{\yng(2,2,1)}\rangle
+(q^{-2}+q^{-3})\langle s_{\yng(2,1,1,1)}\rangle
-(q^{-4}+q^{-5})\langle s_{\yng(1,1,1,1,1)}\rangle
\sim 0.
\end{align}
There are more relations for the case with more boxes.
These relations can be translated into
\begin{align}
&\sigma_{M+1}|\langle s_{\yng(3)}\rangle|
-\sigma_{M}|\langle s_{\yng(2,1)}\rangle|
+\sigma_{M-1}|\langle s_{\yng(1,1,1)}\rangle|=0,\nonumber\\
&(\sigma_{M+2}+\sigma_{M})|\langle s_{\yng(4)}\rangle|
-\sigma_{M}|\langle s_{\yng(3,1)}\rangle|
-\sigma_{M-2}|\langle s_{\yng(2,2)}\rangle|
+\sigma_{M-2}|\langle s_{\yng(2,1,1)}\rangle|=0,\nonumber\\
&\sigma_{M+2}|\langle s_{\yng(3,1)}\rangle|
-\sigma_{M+2}|\langle s_{\yng(2,2)}\rangle|
-\sigma_{M}|\langle s_{\yng(2,1,1)}\rangle|
+(\sigma_{M}+\sigma_{M-2})|\langle s_{\yng(1,1,1,1)}\rangle|=0,
\end{align}
and
\begin{align}
&(\sigma_{M+2}+\sigma_{M})|\langle s_{\yng(5)}\rangle|
-(\sigma_{M+1}+\sigma_{M-1})|\langle s_{\yng(4,1)}\rangle|
+\sigma_{M}|\langle s_{\yng(3,2)}\rangle|
+\sigma_{M-2}|\langle s_{\yng(3,1,1)}\rangle|
-\sigma_{M-2}|\langle s_{\yng(2,2,1)}\rangle|\nonumber\\
&\hspace{15cm}=0,\nonumber\\
&\sigma_{M-1}|\langle s_{\yng(4,1)}\rangle|
-(\sigma_{M}+\sigma_{M-2})|\langle s_{\yng(3,2)}\rangle|
+(\sigma_{M}+\sigma_{M-2})|\langle s_{\yng(2,2,1)}\rangle|
-\sigma_{M-1}|\langle s_{\yng(2,1,1,1)}\rangle|=0,\nonumber\\
&(\sigma_{M+3}+\sigma_{M+1})|\langle s_{\yng(4,1)}\rangle|
-\sigma_{M+2}|\langle s_{\yng(3,2)}\rangle|
-\sigma_{M}|\langle s_{\yng(3,1,1)}\rangle|
-\sigma_{M-2}|\langle s_{\yng(2,2,1)}\rangle|
+(\sigma_{M-1}+\sigma_{M-3})|\langle s_{\yng(2,1,1,1)}\rangle|\nonumber\\
&\hspace{15cm}=0,\nonumber\\
&\sigma_{M+1}|\langle s_{\yng(4,1)}\rangle|
-(\sigma_{M+2}+\sigma_{M})|\langle s_{\yng(3,2)}\rangle|
+(\sigma_{M+2}+\sigma_{M})|\langle s_{\yng(2,2,1)}\rangle|
-\sigma_{M+1}|\langle s_{\yng(2,1,1,1)}\rangle|=0,\nonumber\\
&\sigma_{M+2}|\langle s_{\yng(3,2)}\rangle|
+\sigma_{M+2}|\langle s_{\yng(3,1,1)}\rangle|
+\sigma_{M}|\langle s_{\yng(2,2,1)}\rangle|
-(\sigma_{M+1}+\sigma_{M-1})|\langle s_{\yng(2,1,1,1)}\rangle|
+(\sigma_{M}+\sigma_{M-2})|\langle s_{\yng(1,1,1,1,1)}\rangle|=0.
\end{align}

\subsection{Check of the lowest component}\label{lowcheck}

In this appendix we compute the lowest component for the both sides of the descent relation and find a check for the relation.

We first consider the descent relation \eqref{hookbox} between the hook representation and the fundamental representation, 
\begin{align}
\im\Bigl[e^{-i\Theta^{(a|l),\yng(1)}_{k,M}}\langle s_{(a|l)}\bar s_\Box\rangle
^\text{GC}_{k,M}(z)\Bigr]
=\im\Bigl[e^{-i\Theta^{(a|l),\yng(1)}_{k,M}}
\bigl(\langle s_{(a-1|l)}\rangle^\text{GC}_{k,M}(z)
+\langle s_{(a|l-1)}\rangle^\text{GC}_{k,M}(z)\bigr)\Bigr].
\end{align}
We can compute the lowest component on both sides and find that, when $M=0$ the lowest component on both sides is
\begin{align}
\frac{z}{k}\sin\frac{\pi(-a+l)}{k},
\end{align}
when $1\le M\le a-\frac{1}{2}$ the lowest component is
\begin{align}
\frac{z}{k^{1+\frac{M}{2}}}
\frac{\prod_{j=1}^{M-1}(2\sin\frac{\pi j}{k})^{M-j}
\cdot\prod_{j=l+\frac{1}{2}}^{l+M-\frac{1}{2}}2\sin\frac{\pi j}{k}}
{\prod_{j=a-M+\frac{1}{2}}^{a-\frac{1}{2}}2\cos\frac{\pi j}{k}}\sin\frac{\pi(-a+l+M)}{k},
\label{a-1/2}
\end{align}
and when $a+\frac{1}{2}\le M$ the lowest component is
\begin{align}
\frac{1}{k^{\frac{M}{2}}}\frac{\prod_{j=1}^{M-1}(2\sin\frac{\pi j}{k})^{M-j}
\cdot\prod_{j=l+\frac{1}{2}}^{l+M-\frac{1}{2}}2\sin\frac{\pi j}{k}}
{\prod_{j=1}^{a-\frac{1}{2}}2\sin\frac{\pi j}{k}
\cdot\prod_{j=1}^{M-a-\frac{1}{2}}2\sin\frac{\pi j}{k}}
\cos\frac{\pi(-a+l+M)}{k}.
\end{align}
Note that a special care is needed in discussing the case of $M=a-\frac{1}{2}$.
For this case, both the left-hand side and the second term on the right-hand side contribute as ${\cal O}(z)$ in \eqref{a-1/2} while the first term on the right-hand side does not contribute if we assume that the phase of the one-point function is constant.

Next let us turn to the check of the descent relation \eqref{hooksym} between the hook representation and the symmetric representation
\begin{align}
\im\Bigl[e^{-i\Theta^{(a|\frac{1}{2}),\yng(2)}_{k,M}}
\langle s_{(a|\frac{1}{2})}\bar s_{\yng(2)}\rangle^\text{GC}_{k,M}(z)\Bigr]
=\im\Bigl[e^{-i\Theta^{(a|\frac{1}{2}),\yng(2)}_{k,M}}
(q^{a-\frac{1}{2}}\langle s_{(a|\frac{1}{2})}\rangle^\text{GC}_{k,M}(z)
+q^{-1}\langle s_{(a-1|\frac{3}{2})}\rangle^\text{GC}_{k,M}(z))\Bigr],
\end{align}
with $a>1$.
Again, we can compute the lowest component on both sides and find that, when $M=0$ the lowest component on both sides is
\begin{align}
-\frac{z}{k}\frac{1}
{2\cos\frac{\pi(a+\frac{1}{2})}{k}}
\biggl[\sin\frac{4\pi a}{k}+\sin\frac{2\pi(a-\frac{1}{2})}{k}\biggr],
\end{align}
when $1\le M\le a-\frac{1}{2}$ the lowest component is
\begin{align}
-\frac{z}{k^{1+\frac{M}{2}}}\frac{\prod_{j=1}^M(2\sin\frac{\pi j}{k})^{M+1-j}}
{\prod_{j=a-M+\frac{1}{2}}^{a+\frac{1}{2}}2\cos\frac{\pi j}{k}}
\biggl[\sin\frac{4\pi(a-\frac{M}{2})}{k}
+\frac{\sin\frac{2\pi(a-\frac{M+1}{2})}{k}\sin\frac{\pi(M+1)}{k}}{\sin\frac{\pi}{k}}
-\frac{\sin\frac{\pi M}{k}\sin\frac{\pi(M+1)}{k}}{\sin\frac{\pi}{k}}\biggr],
\end{align}
and when $a+\frac{1}{2}\le M$ the lowest component is
\begin{align}
&-\frac{1}{k^{\frac{M}{2}}}\frac{\prod_{j=1}^M(2\sin\frac{\pi j}{k})^{M+1-j}}
{\prod_{j=1}^{a+\frac{1}{2}}2\sin\frac{\pi j}{k}
\cdot\prod_{j=1}^{M-a-\frac{1}{2}}2\sin\frac{\pi j}{k}}\nonumber\\
&\quad\times\biggl[\sin\frac{4\pi(a-\frac{M}{2})}{k}
-\frac{\sin\frac{2\pi(a-\frac{M+1}{2})}{k}\sin\frac{\pi(M+1)}{k}}{\sin\frac{\pi}{k}}
-\frac{\sin\frac{\pi M}{k}\sin\frac{\pi(M+1)}{k}}{\sin\frac{\pi}{k}}\biggr].
\end{align}
As before a special care is necessary for $M=a-\frac{1}{2}$.

\subsection{Towards interpretation}\label{heisenberg}

Stimulated by the reduction by two boxes in \eqref{imaginary} and the explicit relation \eqref{generalbox}, we attempt an interpretation in the ``operator formalism''.
Since taking the imaginary part on the left-hand side of \eqref{imaginary} amounts to the subtraction by its conjugation, we are naturally led to the study of the ``Poisson bracket'' (without the anticommutativity).
To be concrete, we define the bracket by
\begin{align}
\{s_Y,s_Z\}=\lim_{N\to\infty}\frac{1}{N}\sum_{n=1}^N
\frac{\partial s_Y(x)}{\partial x_n}\frac{\partial s_Z(x)}{\partial x_n}.
\label{bracket}
\end{align}
Then, it is not difficult to check the relation
\begin{align}
\{s_Y,s_\Box\}=\sum_{Y_\bullet}s_{Y_\bullet},
\end{align}
up to total boxes of six which resembles \eqref{generalbox}.
Aside from the pattern, we also find
\begin{align}
&\{s_{\yng(2)},s_{\yng(2)}\}
=s_{\yng(2)}+s_{\yng(1,1)},\quad
\{s_{\yng(1,1)},s_{\yng(2)}\}
=s_{\yng(2)}+s_{\yng(1,1)},\quad
\{s_{\yng(1,1)},s_{\yng(1,1)}\}
=s_{\yng(2)}+s_{\yng(1,1)},\nonumber\\
&\{s_{\yng(3)},s_{\yng(2)}\}
=s_{\yng(3)}+s_{\yng(2,1)},\quad
\{s_{\yng(2,1)},s_{\yng(2)}\}
=s_{\yng(3)}+2s_{\yng(2,1)}+s_{\yng(1,1,1)},\quad
\{s_{\yng(1,1,1)},s_{\yng(2)}\}
=s_{\yng(2,1)}+s_{\yng(1,1,1)},\nonumber\\
&\{s_{\yng(3)},s_{\yng(1,1)}\}
=s_{\yng(3)}+s_{\yng(2,1)},\quad
\{s_{\yng(2,1)},s_{\yng(1,1)}\}
=s_{\yng(3)}+2s_{\yng(2,1)}+s_{\yng(1,1,1)},\quad
\{s_{\yng(1,1,1)},s_{\yng(1,1)}\}
=s_{\yng(2,1)}+s_{\yng(1,1,1)},\nonumber\\
&\{s_{\yng(4)},s_{\yng(2)}\}
=s_{\yng(4)}+s_{\yng(3,1)},\quad
\{s_{\yng(3,1)},s_{\yng(2)}\}
=s_{\yng(4)}+2s_{\yng(3,1)}+s_{\yng(2,2)}+s_{\yng(2,1,1)},\nonumber\\
&\{s_{\yng(2,2)},s_{\yng(2)}\}
=s_{\yng(3,1)}+s_{\yng(2,2)}+s_{\yng(2,1,1)},\quad
\{s_{\yng(2,1,1)},s_{\yng(2)}\}
=s_{\yng(3,1)}+s_{\yng(2,2)}+2s_{\yng(2,1,1)}+s_{\yng(1,1,1,1)},\quad
\{s_{\yng(1,1,1,1)},s_{\yng(2)}\}
=s_{\yng(2,1,1)}+s_{\yng(1,1,1,1)},\nonumber\\
&\{s_{\yng(4)},s_{\yng(1,1)}\}
=s_{\yng(4)}+s_{\yng(3,1)},\quad
\{s_{\yng(3,1)},s_{\yng(1,1)}\}
=s_{\yng(4)}+2s_{\yng(3,1)}+s_{\yng(2,2)}+s_{\yng(2,1,1)},\nonumber\\
&\{s_{\yng(2,2)},s_{\yng(1,1)}\}
=s_{\yng(3,1)}+s_{\yng(2,2)}+s_{\yng(2,1,1)},\quad
\{s_{\yng(2,1,1)},s_{\yng(1,1)}\}
=s_{\yng(3,1)}+s_{\yng(2,2)}+2s_{\yng(2,1,1)}+s_{\yng(1,1,1,1)},\quad
\{s_{\yng(1,1,1,1)},s_{\yng(1,1)}\}
=s_{\yng(2,1,1)}+s_{\yng(1,1,1,1)},
\label{2&11}
\end{align}
for $Z=\yng(2)$ and $Z=\yng(1,1)$ and
\begin{align}
&\{s_{\yng(3)},s_{\yng(3)}\}
=s_{\yng(4)}+s_{\yng(3,1)}+s_{\yng(2,2)},\quad
\{s_{\yng(2,1)},s_{\yng(3)}\}
=s_{\yng(4)}+2s_{\yng(3,1)}+s_{\yng(2,2)}+s_{\yng(2,1,1)},\nonumber\\
&\{s_{\yng(1,1,1)},s_{\yng(3)}\}
=s_{\yng(3,1)}+s_{\yng(2,1,1)},\quad
\{s_{\yng(2,1)},s_{\yng(2,1)}\}
=s_{\yng(4)}+3s_{\yng(3,1)}+2s_{\yng(2,2)}+3s_{\yng(2,1,1)}+s_{\yng(1,1,1,1)},\nonumber\\
&\{s_{\yng(1,1,1)},s_{\yng(2,1)}\}
=s_{\yng(3,1)}+s_{\yng(2,2)}+2s_{\yng(2,1,1)}+s_{\yng(1,1,1,1)},\quad
\{s_{\yng(1,1,1)},s_{\yng(1,1,1)}\}
=s_{\yng(2,2)}+s_{\yng(2,1,1)}+s_{\yng(1,1,1,1)},
\label{Z3boxes}
\end{align}
for $|Z|=3$.
Most of the results closely resemble the descent relation we have found in appendix \ref{interference} if we set all of $q=e^{-\frac{4\pi i}{k}}$ to be $1$.
There are several exceptions.
For example, in the case of five total boxes \eqref{5descent}, after setting $q\to 1$, both $\langle s_{\yng(2,1)}\bar s_{\yng(2)}\rangle$ and $\langle s_{\yng(2,1)}\bar s_{\yng(1,1)}\rangle$ reduce to $3\langle s_{\yng(2,1)}\rangle$ instead of $\langle s_{\yng(3)}\rangle+2\langle s_{\yng(2,1)}\rangle+\langle s_{\yng(1,1,1)}\rangle$ which is expected from \eqref{2&11}.
So naturally our next question would be whether we can use the ambiguities discussed in appendix \ref{ambiguity} to improve the descent relation in appendix \ref{interference} so that the result of the bracket \eqref{bracket} is correctly reproduced in the limit $q\to 1$.
We have found that, up to six total boxes, the answer is yes at the price of allowing half-integral coefficients.\footnote{We can alternatively allow half-integral powers of $q$ while keeping integral coefficients.}
After the improvements, the relations read
\begin{align}
&2\langle s_{\yng(2,1)}\bar s_{\yng(2)}\rangle
\sim(q+1)\langle s_{\yng(3)}\rangle
+(2q+1+q^{-1})\langle s_{\yng(2,1)}\rangle
+(q^{-1}+q^{-2})\langle s_{\yng(1,1,1)}\rangle,\nonumber\\
&2\langle s_{\yng(2,1)}\bar s_{\yng(1,1)}\rangle
\sim(q^2+q)\langle s_{\yng(3)}\rangle
+(q+1+2q^{-1})\langle s_{\yng(2,1)}\rangle
+(1+q^{-1})\langle s_{\yng(1,1,1)}\rangle,\nonumber\\
&2\langle s_{\yng(3,1)}\bar s_{\yng(2)}\rangle
\sim(q^2+1)\langle s_{\yng(4)}\rangle
+(2q^2+1+q^{-1})\langle s_{\yng(3,1)}\rangle
+(1+q^{-1})\langle s_{\yng(2,2)}\rangle
+(q^{-1}+q^{-2})\langle s_{\yng(2,1,1)}\rangle,\nonumber\\
&2\langle s_{\yng(2,2)}\bar s_{\yng(2)}\rangle
\sim q(q-1)^2\langle s_{\yng(4)}\rangle
+(q+1)\langle s_{\yng(3,1)}\rangle
+(q+3-2q^{-1})\langle s_{\yng(2,2)}\rangle
+(1-q^{-1}+2q^{-2})\langle s_{\yng(2,1,1)}\rangle,\nonumber\\
&2\langle s_{\yng(2,1,1)}\bar s_{\yng(2)}\rangle
\sim(q+1)\langle s_{\yng(3,1)}\rangle
+(1+q^{-1})\langle s_{\yng(2,2)}\rangle
+(2q+1+q^{-1})\langle s_{\yng(2,1,1)}\rangle
+(q^{-1}+q^{-3})\langle s_{\yng(1,1,1,1)}\rangle,
\nonumber\\
&2\langle s_{\yng(3,1)}\bar s_{\yng(1,1)}\rangle
\sim(q^3+q)\langle s_{\yng(4)}\rangle
+(q+1+2q^{-1})\langle s_{\yng(3,1)}\rangle
+(q+1)\langle s_{\yng(2,2)}\rangle
+(1+q^{-1})\langle s_{\yng(2,1,1)}\rangle,\nonumber\\
&2\langle s_{\yng(2,2)}\bar s_{\yng(1,1)}\rangle
\sim(2q^2-q+1)\langle s_{\yng(3,1)}\rangle
+(-2q+3+q^{-1})\langle s_{\yng(2,2)}\rangle
+(1+q^{-1})\langle s_{\yng(2,1,1)}\rangle
+q^{-1}(1-q^{-1})^2\langle s_{\yng(1,1,1,1)}\rangle,\nonumber\\
&2\langle s_{\yng(2,1,1)}\bar s_{\yng(1,1)}\rangle
\sim(q^2+q)\langle s_{\yng(3,1)}\rangle
+(q+1)\langle s_{\yng(2,2)}\rangle
+(q+1+2q^{-2})\langle s_{\yng(2,1,1)}\rangle
+(1+q^{-2})\langle s_{\yng(1,1,1,1)}\rangle,\nonumber\\
&2\langle s_{\yng(2,1)}\bar s_{\yng(2,1)}\rangle
\sim(q^3+q^2)\langle s_{\yng(4)}\rangle
+(2q^2-q+5)\langle s_{\yng(3,1)}\rangle
+(2q^2-q+2-q^{-1}+2q^{-2})\langle s_{\yng(2,2)}\rangle
\nonumber\\
&\qquad
+(5-q^{-1}+2q^{-2})\langle s_{\yng(2,1,1)}\rangle
+(q^{-2}+q^{-3})\langle s_{\yng(1,1,1,1)}\rangle.
\label{improvements}
\end{align}
Note that since the descent relations \eqref{descentYZ42}, \eqref{descentYZ33} and the brackets \eqref{2&11}, \eqref{Z3boxes} match both between $\langle s_{\yng(3,1)}\bar s_{\yng(2)}\rangle$ and $\langle s_{\yng(2,1)}\bar s_{\yng(3)}\rangle$ and between $\langle s_{\yng(2,1,1)}\bar s_{\yng(1,1)}\rangle$ and $\langle s_{\yng(1,1,1)}\bar s_{\yng(2,1)}\rangle$, the improvements \eqref{improvements} also match and we omit the latter cases.

\section{Relations among one-point functions}\label{onept}

In this appendix we shall list some relations among the one-point functions discussed in section \ref{oneptrel}, which are obtained by combining two relations for the two-point functions.
We present the result with the notation introduced in section \ref{oneptrel}.
For the case of less than four boxes we have
\begin{align}
&\sigma_1|\langle s_{\yng(2)}\rangle|
-\sigma_1|\langle s_{\yng(1,1)}\rangle|
+\sigma_{M}|\langle 1\rangle|=0,\nonumber\\
&\sigma_2|\langle s_{\yng(3)}\rangle|
-\sigma_1|\langle s_{\yng(2,1)}\rangle|
+\sigma_{M-1}|\langle s_{\yng(1)}\rangle|=0,\nonumber\\
&\sigma_1|\langle s_{\yng(2,1)}\rangle|
-\sigma_2|\langle s_{\yng(1,1,1)}\rangle|
+\sigma_{M+1}|\langle s_{\yng(1)}\rangle|=0,
\end{align}
for the case of four boxes we have
\begin{align}
&\sigma_3|\langle s_{\yng(4)}\rangle|
-\sigma_1|\langle s_{\yng(3,1)}\rangle|
+\sigma_{M-2}|\langle s_{\yng(2)}\rangle|=0,\nonumber\\
&\sigma_4|\langle s_{\yng(4)}\rangle|
-\sigma_2|\langle s_{\yng(2,2)}\rangle|
+\sigma_{M-3}|\langle s_{\yng(2)}\rangle|
+\sigma_{M-1}|\langle s_{\yng(1,1)}\rangle|=0,\nonumber\\
&\sigma_2|\langle s_{\yng(3,1)}\rangle|
-\sigma_2|\langle s_{\yng(2,1,1)}\rangle|
+\sigma_{M+1}|\langle s_{\yng(2)}\rangle|
+\sigma_{M-1}|\langle s_{\yng(1,1)}\rangle|=0,\nonumber\\
&\sigma_2|\langle s_{\yng(2,2)}\rangle|
-\sigma_4|\langle s_{\yng(1,1,1,1)}\rangle|
+\sigma_{M+1}|\langle s_{\yng(2)}\rangle|
+\sigma_{M+3}|\langle s_{\yng(1,1)}\rangle|=0,\nonumber\\
&\sigma_1|\langle s_{\yng(2,1,1)}\rangle|
-\sigma_3|\langle s_{\yng(1,1,1,1)}\rangle|
+\sigma_{M+2}|\langle s_{\yng(1,1)}\rangle|=0,
\end{align}
and for the case of five boxes we have
\begin{align}
&\sigma_{4}|\langle s_{\yng(5)}\rangle|
-\sigma_{1}|\langle s_{\yng(4,1)}\rangle|
+\sigma_{M-3}|\langle s_{\yng(3)}\rangle|=0,\nonumber\\
&\sigma_{6}|\langle s_{\yng(5)}\rangle|
+\sigma_{1}|\langle s_{\yng(4,1)}\rangle|
-\sigma_{2}|\langle s_{\yng(3,2)}\rangle|
+\sigma_{M-5}|\langle s_{\yng(3)}\rangle|
+\sigma_{M-2}|\langle s_{\yng(2,1)}\rangle|=0,\nonumber\\
&\sigma_{3}|\langle s_{\yng(4,1)}\rangle|
-\sigma_{2}|\langle s_{\yng(3,1,1)}\rangle|
+\sigma_{M+1}|\langle s_{\yng(3)}\rangle|
+\sigma_{M-2}|\langle s_{\yng(2,1)}\rangle|=0,\nonumber\\
&\sigma_{4}|\langle s_{\yng(4,1)}\rangle|
+\sigma_{1}|\langle s_{\yng(3,2)}\rangle|
-\sigma_{1}|\langle s_{\yng(3,1,1)}\rangle|
-\sigma_{3}|\langle s_{\yng(2,2,1)}\rangle|
+(\sigma_{M-3}+\sigma_{M-1}+\sigma_{M+1})|\langle s_{\yng(2,1)}\rangle|=0,\nonumber\\
&\sigma_{2}|\langle s_{\yng(3,2)}\rangle|
-\sigma_{2}|\langle s_{\yng(2,2,1)}\rangle|
+\sigma_{M}|\langle s_{\yng(2,1)}\rangle|=0,\nonumber\\
&\sigma_{3}|\langle s_{\yng(3,2)}\rangle|
+\sigma_{1}|\langle s_{\yng(3,1,1)}\rangle|
-\sigma_{1}|\langle s_{\yng(2,2,1)}\rangle|
-\sigma_{4}|\langle s_{\yng(2,1,1,1)}\rangle|
+(\sigma_{M+3}+\sigma_{M+1}+\sigma_{M-1})|\langle s_{\yng(2,1)}\rangle|=0,\nonumber\\
&\sigma_{2}|\langle s_{\yng(3,1,1)}\rangle|
-\sigma_{3}|\langle s_{\yng(2,1,1,1)}\rangle|
+\sigma_{M+2}|\langle s_{\yng(2,1)}\rangle|
+\sigma_{M-1}|\langle s_{\yng(1,1,1)}\rangle|=0,\nonumber\\
&\sigma_{2}|\langle s_{\yng(2,2,1)}\rangle|
-\sigma_{1}|\langle s_{\yng(2,1,1,1)}\rangle|
-\sigma_{6}|\langle s_{\yng(1,1,1,1,1)}\rangle|
+\sigma_{M+2}|\langle s_{\yng(2,1)}\rangle|
+\sigma_{M+5}|\langle s_{\yng(1,1,1)}\rangle|=0,\nonumber\\
&\sigma_{1}|\langle s_{\yng(2,1,1,1)}\rangle|
-\sigma_{4}|\langle s_{\yng(1,1,1,1,1)}\rangle|
+\sigma_{M+3}|\langle s_{\yng(1,1,1)}\rangle|=0.
\end{align}

\section{Perturbative one-point function}\label{pert1pt}

\subsection{Giambelli compatibility}\label{giambellic}

In this appendix we show that the perturbative part of the one-point function \eqref{1ptpert} satisfies the Giambelli identity \eqref{Giambelliid}.
Since the numerators which count the box number of the Young diagram $|Y|$ or add up the contents $c^Y$ \eqref{contents} cancel trivially on both sides, the proof reduces to that of the denominators
\begin{align}
\prod_{x\in(a_1,\cdots,a_R|l_1,\cdots,l_R)}\frac{1}{2\sin\frac{2\pi h(x)}{k}}
=\det\biggl(\prod_{x\in(a_i|l_j)}\frac{1}{2\sin\frac{2\pi h(x)}{k}}\biggr)
_{\begin{subarray}{c}1\le i\le R\\1\le j\le R\end{subarray}}.
\label{qdim}
\end{align}
This is because the original Schur polynomials $s_Y(x)$ satisfy the Giambelli identity (see Section I.3, Example 9 in \cite{Macdonald})
\begin{align}
s_{(a_1,\cdots,a_R|l_1,\cdots,l_R)}(x_1,\cdots,x_n)
=\det\bigl(s_{(a_i|l_j)}(x_1,\cdots,x_n)\bigr)
_{\begin{subarray}{c}1\le i\le R\\1\le j\le R\end{subarray}},
\end{align}
and after the specification $x_i=q^{i-\frac{1}{2}}$ the Schur polynomials reduce to (see Section I.3, Example 1 in \cite{Macdonald} combined with Section I.1, Examples 2 and 3 to eliminate some unwanted quantities)
\begin{align}
s_Y(q^{\frac{1}{2}},q^{\frac{3}{2}},\cdots,q^{n-\frac{1}{2}})
=\prod_{x\in Y}\frac{q^{\frac{1}{2}h(x)}}{1-q^{h(x)}}\frac{1-q^{n+c(x)}}{q^{\frac{1}{2}c(x)}},
\label{schurpower}
\end{align}
where $c(x)$ denotes the content at $x\in Y$, $c(i,j)=j-i$.
Since the factor unrelated the hook length cancels on both sides in the limit $n\to\infty$, the only remaining factor is
\begin{align}
\prod_{x\in(a_1,\cdots,a_R|l_1,\cdots,l_R)}\frac{q^{\frac{1}{2}h(x)}}{1-q^{h(x)}}
=\det\biggl(\prod_{x\in(a_i|l_j)}\frac{q^{\frac{1}{2}h(x)}}{1-q^{h(x)}}\biggr)
_{\begin{subarray}{c}1\le i\le R\\1\le j\le R\end{subarray}}.
\end{align}
After substituting $q=e^{-\frac{4\pi i}{k}}$, the formula reduces to \eqref{qdim}.

\subsection{Free energy of open topological strings}\label{freeopen}

In this appendix we show that, with the correspondence \eqref{mattop}, the perturbative part of the one-point function \eqref{1ptpert} can be derived from the free energy of the open topological string theory \eqref{openfree} with the identification \eqref{Vhat}.
If the only contribution for $(d_+,d_-)=(0,0)$ comes from $n^{{\bm d}={\bm 0},{\bm\ell}=(1)}_{g=0}=1$, the free energy of the open topological string theory is given by
\begin{align}
F^\text{op}(\widehat V)
=\sum_{n=1}^\infty\frac{1}{n}\frac{1}{2i\sin\frac{2\pi n}{k}}\tr\widehat V^n.
\end{align}
The sine function in the denominator can be expanded as
\begin{align}
F^\text{op}(\widehat V)
=\sum_{n=1}^\infty\sum_{m=0}^\infty\frac{1}{n}\tr\Bigl[\widehat Vq^{m+\frac{1}{2}}\Bigr]^n,
\end{align}
with $q=e^{-\frac{4\pi i}{k}}$.
It is enough to only consider the diagonal matrix $\widehat V=\diag(\widehat V_i)$. 
If we sum over $n$ first, the free energy becomes
\begin{align}
F^\text{op}(\widehat V)=\sum_i\sum_{m=0}^\infty-\log\Bigl(1-\widehat V_iq^{m+\frac{1}{2}}\Bigr),
\end{align}
which, after the exponentiation, implies
\begin{align}
e^{F^\text{op}(\widehat V)}
=\sum_Ys_Y(q^{\frac{1}{2}},q^{\frac{3}{2}},\cdots)\tr_Y(\widehat V),
\end{align}
where we have used the orthogonal relation
\begin{align}
\prod_{i,j}\frac{1}{1-x_iy_j}=\sum_Ys_Y(x)s_Y(y).
\end{align}
Using \eqref{schurpower}, we can further rewrite the free energy of the open topological string theory as
\begin{align}
e^{F^\text{op}(\widehat V)}
=\sum_Y\Biggl[\prod_{x\in Y}\frac{e^{\frac{2\pi i}{k}c(x)}}{2\sin\frac{2\pi h(x)}{k}}\Biggr]
e^{|Y|(\frac{2\mu}{k}-\pi i\frac{M}{k})}\tr_Y(V),
\end{align}
which takes care of both the phase factor \eqref{1ptphase} by
\begin{align}
\Biggl[\prod_{x\in Y}e^{\frac{2\pi i}{k}c(x)}\Biggr]e^{-\frac{\pi i}{k}M|Y|}=e^{i\theta_{k,M}^Y},
\end{align}
and the remaining perturbative part \eqref{1ptpert}.

\section{Non-perturbative one-point function}\label{np1pt}

In this appendix, we list some values of the one-point function in the grand canonical ensemble $\langle s_Y\rangle^\text{GC}_{k,M}$ obtained numerically to study the relation to the free energy of the open topological string theory.
Since the phase factor of the one-point function is trivial, we list the reduced one-point function in the grand canonical ensemble with the perturbative part removed
\begin{align}
\W^Y_{k,M}
=\frac{\llangle s_Y\rrangle^\text{GC}_{k,M}}{\llangle 1\rrangle^\text{GC}_{k,M}}
\Bigg/\frac{\langle s_Y\rangle^\text{GC}_{k,M}}
{\langle 1\rangle^\text{GC}_{k,M}}\Bigg|^\text{pert},
\end{align}
in terms of $Q=-e^{-\frac{4\mu_\text{eff}}{k}}$.

For the case that the total box number is one, $|Y|=1$, there is only the fundamental representation $\W_{k,M}^{\yng(1)}$.
For the convergent combination of $(k,M)$, the values of $\W_{k,M}^{\yng(1)}$ are given by
\begin{align}
\W_{3,0}^{\yng(1)}
&=1+2Q+3Q^2+10Q^3+25Q^4+54Q^5+143Q^6+364Q^7+{\cal O}(Q^8),
\nonumber\\
\W_{4,0}^{\yng(1)}
&=1+2Q+3Q^2+10Q^3+17Q^4+64Q^5+132Q^6+494Q^7+{\cal O}(Q^8),
\nonumber\\
\W_{4,1}^{\yng(1)}
&=1+3Q^2-11Q^4+64Q^6+{\cal O}(Q^8),
\nonumber\\
\W_{6,0}^{\yng(1)}
&=1+2Q+3Q^2+10Q^3+25Q^4+54Q^5+143Q^6+364Q^7+{\cal O}(Q^8),
\nonumber\\
\W_{6,1}^{\yng(1)}
&=1+Q+3Q^2+5Q^3+4Q^4-10Q^6-10Q^7+{\cal O}(Q^8),
\nonumber\\
\W_{6,2}^{\yng(1)}
&=1-Q+3Q^2-5Q^3+4Q^4-10Q^6+10Q^7+{\cal O}(Q^8),
\nonumber\\
\W_{8,0}^{\yng(1)}
&=1+2Q+3Q^2+10Q^3+33Q^4+88Q^5+228Q^6+646Q^7+{\cal O}(Q^8),
\nonumber\\
\W_{8,1}^{\yng(1)}
&=1+\sqrt{2}Q+3Q^2+5\sqrt{2}Q^3+19Q^4+26\sqrt{2}Q^5+78Q^6+117\sqrt{2}Q^7+{\cal O}(Q^8),
\nonumber\\
\W_{8,2}^{\yng(1)}
&=1+3Q^2+5Q^4+16Q^6+{\cal O}(Q^8),
\nonumber\\
\W_{8,3}^{\yng(1)}
&=1-\sqrt{2}Q+3Q^2-5\sqrt{2}Q^3+19Q^4-26\sqrt{2}Q^5+78Q^6-117\sqrt{2}Q^7+{\cal O}(Q^8),
\nonumber\\
\W_{12,0}^{\yng(1)}
&=1+2Q+3Q^2+10Q^3+41Q^4+166Q^5+615Q^6+2156Q^7+{\cal O}(Q^8),
\nonumber\\
\W_{12,1}^{\yng(1)}
&=1+\sqrt{3}Q+3Q^2+5\sqrt{3}Q^3+34Q^4+74\sqrt{3}Q^5+442Q^6+840\sqrt{3}Q^7+{\cal O}(Q^8),
\nonumber\\
\W_{12,2}^{\yng(1)}
&=1+Q+3Q^2+5Q^3+20Q^4+56Q^5+162Q^6+442Q^7+{\cal O}(Q^8),
\nonumber\\
\W_{12,3}^{\yng(1)}
&=1+3Q^2+13Q^4+55Q^6+{\cal O}(Q^8),
\nonumber\\
\W_{12,4}^{\yng(1)}
&=1-Q+3Q^2-5Q^3+20Q^4-56Q^5+162Q^6-442Q^7+{\cal O}(Q^8),
\nonumber\\
\W_{12,5}^{\yng(1)}
&=1-\sqrt{3}Q+3Q^2-5\sqrt{3}Q^3+34Q^4-74\sqrt{3}Q^5+442Q^6-840\sqrt{3}Q^7+{\cal O}(Q^8).
\end{align}
For the case that the total box number is two, $|Y|=2$, we have the symmetric representation and the anti-symmetric representation.
For the convergent combinations of $(k,M)$, the one-point function in the symmetric representation $\W_{k,M}^{\yng(2)}$ is given by
\begin{align}
\W_{6,0}^{\yng(2)}
&=1+Q+4Q^2+8Q^3+18Q^4+52Q^5+116Q^6+288Q^7+{\cal O}(Q^8),
\nonumber\\
\W_{6,1}^{\yng(2)}
&=1+2Q+Q^2+4Q^3+2Q^5-13Q^6+{\cal O}(Q^8),
\nonumber\\
\W_{6,2}^{\yng(2)}
&=1+Q+Q^2-4Q^3-2Q^5-13Q^6+{\cal O}(Q^8),
\nonumber\\
\W_{8,0}^{\yng(2)}
&=1+2Q+6Q^2+16Q^3+47Q^4+128Q^5+358Q^6+992Q^7+{\cal O}(Q^8),
\nonumber\\
\W_{8,1}^{\yng(2)}
&=1+2\sqrt{2}Q+4Q^2+8\sqrt{2}Q^3+23Q^4+32\sqrt{2}Q^5+100Q^6+144\sqrt{2}Q^7+{\cal O}(Q^8),
\nonumber\\
\W_{8,2}^{\yng(2)}
&=1+2Q+2Q^2-Q^4+2Q^6+{\cal O}(Q^8),
\nonumber\\
\W_{8,3}^{\yng(2)}
&=1+4Q^2-8\sqrt{2}Q^3+23Q^4-32\sqrt{2}Q^5+100Q^6-144\sqrt{2}Q^7+{\cal O}(Q^8),
\nonumber\\
\W_{12,0}^{\yng(2)}
&=1+3Q+8Q^2+24Q^3+90Q^4+348Q^5+1288Q^6+4608Q^7+{\cal O}(Q^8),
\nonumber\\
\W_{12,1}^{\yng(2)}
&=1+2\sqrt{3}Q+7Q^2+12\sqrt{3}Q^3+72Q^4+150\sqrt{3}Q^5+893Q^6+1728\sqrt{3}Q^7+{\cal O}(Q^8),
\nonumber\\
\W_{12,2}^{\yng(2)}
&=1+3Q+5Q^2+12Q^3+36Q^4+102Q^5+283Q^6+792Q^7+{\cal O}(Q^8),
\nonumber\\
\W_{12,3}^{\yng(2)}
&=1+\sqrt{3}Q+4Q^2+18Q^4+68Q^6+{\cal O}(Q^8),
\nonumber\\
\W_{12,4}^{\yng(2)}
&=1+5Q^2-12Q^3+36Q^4-102Q^5+283Q^6-792Q^7+{\cal O}(Q^8),
\nonumber\\
\W_{12,5}^{\yng(2)}
&=1-\sqrt{3}Q+7Q^2-12\sqrt{3}Q^3+72Q^4-150\sqrt{3}Q^5+893Q^6-1728\sqrt{3}Q^7+{\cal O}(Q^8),
\end{align}
while the one-point function in the anti-symmetric representation $\W_{k,M}^{\yng(1,1)}$ is given by
\begin{align}
\W_{6,0}^{\yng(1,1)}
&=1+Q+4Q^2+8Q^3+18Q^4+52Q^5+116Q^6+288Q^7+{\cal O}(Q^8),
\nonumber\\
\W_{6,1}^{\yng(1,1)}
&=1-Q+Q^2+4Q^3+2Q^5-13Q^6+{\cal O}(Q^8),
\nonumber\\
\W_{8,0}^{\yng(1,1)}
&=1+2Q+6Q^2+16Q^3+47Q^4+128Q^5+358Q^6+992Q^7+{\cal O}(Q^8),
\nonumber\\
\W_{8,1}^{\yng(1,1)}
&=1+4Q^2+8\sqrt{2}Q^3+23Q^4+32\sqrt{2}Q^5+100Q^6+144\sqrt{2}Q^7+{\cal O}(Q^8),
\nonumber\\
\W_{8,2}^{\yng(1,1)}
&=1-2Q+2Q^2-Q^4+2Q^6+{\cal O}(Q^8),
\nonumber\\
\W_{12,0}^{\yng(1,1)}
&=1+3Q+8Q^2+24Q^3+90Q^4+348Q^5+1288Q^6+4608Q^7+{\cal O}(Q^8),
\nonumber\\
\W_{12,1}^{\yng(1,1)}
&=1+\sqrt{3}Q+7Q^2+12\sqrt{3}Q^3+72Q^4+150\sqrt{3}Q^5+893Q^6+1728\sqrt{3}Q^7+{\cal O}(Q^8),
\nonumber\\
\W_{12,2}^{\yng(1,1)}
&=1+5Q^2+12Q^3+36Q^4+102Q^5+283Q^6+792Q^7+{\cal O}(Q^8),
\nonumber\\
\W_{12,3}^{\yng(1,1)}
&=1-\sqrt{3}Q+4Q^2+18Q^4+68Q^6+{\cal O}(Q^8),
\nonumber\\
\W_{12,4}^{\yng(1,1)}
&=1-3Q+5Q^2-12Q^3+36Q^4-102Q^5+283Q^6-792Q^7+{\cal O}(Q^8).
\end{align}
In the case that the total box number is three, there are the totally symmetric representation, the mixed representation and the totally anti-symmetric representation.
The convergent combinations of $\W_{k,M}^{\yng(3)}$ are
\begin{align}
\W_{8,0}^{\yng(3)}
&=1+Q^2+8Q^3+18Q^4+40Q^5+117Q^6+360Q^7+{\cal O}(Q^8),
\nonumber\\
\W_{8,1}^{\yng(3)}
&=1+\sqrt{2}Q+3Q^2+5\sqrt{2}Q^3+18Q^4+25\sqrt{2}Q^5+75Q^6+112\sqrt{2}Q^7+{\cal O}(Q^8),
\nonumber\\
\W_{8,2}^{\yng(3)}
&=1+2Q+Q^2+6Q^3-2Q^4+10Q^5+Q^6+32Q^7+{\cal O}(Q^8),
\nonumber\\
\W_{8,3}^{\yng(3)}
&=1+\sqrt{2}Q-Q^2+\sqrt{2}Q^3-2Q^4+13\sqrt{2}Q^5-29Q^6+44\sqrt{2}Q^7+{\cal O}(Q^8),
\nonumber\\
\W_{12,0}^{\yng(3)}
&=1+2Q+8Q^2+32Q^3+116Q^4+426Q^5+1535Q^6+5502Q^7+{\cal O}(Q^8),
\nonumber\\
\W_{12,1}^{\yng(3)}
&=1+2\sqrt{3}Q+10Q^2+18\sqrt{3}Q^3+106Q^4+212\sqrt{3}Q^5+1255Q^6+2444\sqrt{3}Q^7+{\cal O}(Q^8),
\nonumber\\
\W_{12,2}^{\yng(3)}
&=1+4Q+8Q^2+22Q^3+56Q^4+168Q^5+467Q^6+1320Q^7+{\cal O}(Q^8),
\nonumber\\
\W_{12,3}^{\yng(3)}
&=1+2\sqrt{3}Q+4Q^2+6\sqrt{3}Q^3+16Q^4+26\sqrt{3}Q^5+67Q^6+110\sqrt{3}Q^7+{\cal O}(Q^8),
\nonumber\\
\W_{12,4}^{\yng(3)}
&=1+2Q+2Q^2-4Q^3+26Q^4-48Q^5+131Q^6-348Q^7+{\cal O}(Q^8),
\nonumber\\
\W_{12,5}^{\yng(3)}
&=1+4Q^2-12\sqrt{3}Q^3+76Q^4-144\sqrt{3}Q^5+811Q^6-1560\sqrt{3}Q^7+{\cal O}(Q^8),
\end{align}
the convergent combinations of $\W_{k,M}^{\yng(2,1)}$ are
\begin{align}
\W_{8,0}^{\yng(2,1)}
&=1+2Q+5Q^2+14Q^3+38Q^4+106Q^5+293Q^6+816Q^7+{\cal O}(Q^8),
\nonumber\\
\W_{8,1}^{\yng(2,1)}
&=1+\sqrt{2}Q+3Q^2+5\sqrt{2}Q^3+18Q^4+25\sqrt{2}Q^5+75Q^6+112\sqrt{2}Q^7+{\cal O}(Q^8),
\nonumber\\
\W_{8,2}^{\yng(2,1)}
&=1+Q^2-2Q^4+Q^6+{\cal O}(Q^8),
\nonumber\\
\W_{12,0}^{\yng(2,1)}
&=1+4Q+12Q^2+38Q^3+136Q^4+508Q^5+1867Q^6+6732Q^7+{\cal O}(Q^8),
\nonumber\\
\W_{12,1}^{\yng(2,1)}
&=1+2\sqrt{3}Q+10Q^2+18\sqrt{3}Q^3+106Q^4+212\sqrt{3}Q^5+1255Q^6+2444\sqrt{3}Q^7+{\cal O}(Q^8),
\nonumber\\
\W_{12,2}^{\yng(2,1)}
&=1+2Q+6Q^2+16Q^3+46Q^4+128Q^5+355Q^6+996Q^7+{\cal O}(Q^8),
\nonumber\\
\W_{12,3}^{\yng(2,1)}
&=1+4Q^2+16Q^4+67Q^6+{\cal O}(Q^8),
\nonumber\\
\W_{12,4}^{\yng(2,1)}
&=1-2Q+6Q^2-16Q^3+46Q^4-128Q^5+355Q^6-996Q^7+{\cal O}(Q^8),
\end{align}
and the convergent combinations of $\W_{k,M}^{\yng(1,1,1)}$ are
\begin{align}
\W_{8,0}^{\yng(1,1,1)}
&=1+Q^2+8Q^3+18Q^4+40Q^5+117Q^6+360Q^7+{\cal O}(Q^8),
\nonumber\\
\W_{8,1}^{\yng(1,1,1)}
&=1-\sqrt{2}Q-Q^2-\sqrt{2}Q^3-2Q^4-13\sqrt{2}Q^5-29Q^6-44\sqrt{2}Q^7+{\cal O}(Q^8),
\nonumber\\
\W_{12,0}^{\yng(1,1,1)}
&=1+2Q+8Q^2+32Q^3+116Q^4+426Q^5+1535Q^6+5502Q^7+{\cal O}(Q^8),
\nonumber\\
\W_{12,1}^{\yng(1,1,1)}
&=1+4Q^2+12\sqrt{3}Q^3+76Q^4+144\sqrt{3}Q^5+811Q^6+1560\sqrt{3}Q^7+{\cal O}(Q^8),
\nonumber\\
\W_{12,2}^{\yng(1,1,1)}
&=1-2Q+2Q^2+4Q^3+26Q^4+48Q^5+131Q^6+348Q^7+{\cal O}(Q^8),
\nonumber\\
\W_{12,3}^{\yng(1,1,1)}
&=1-2\sqrt{3}Q+4Q^2-6\sqrt{3}Q^3+16Q^4-26\sqrt{3}Q^5+67Q^6-110\sqrt{3}Q^7+{\cal O}(Q^8).
\end{align}
The numerical data in this appendix is helpful in determining the BPS indices in tables \ref{Y1}, \ref{Y11}, \ref{Y2}, \ref{3boxes}.

\section*{Acknowledgements}
We are grateful to Nadav Drukker, Tomohiro Furukawa, Hiroshi Kunitomo, Sho Matsumoto, Tomoki Nosaka, Tadashi Okazaki, Kazuma Shimizu, Kazuhiro Sakai, Seiji Terashima, Koji Umemoto, Kento Watanabe, Yasuhiko Yamada and Katsuya Yano for valuable discussions and/or previous stimulating conversations.
The work of S.M. is supported by JSPS Grant-in-Aid for Scientific Research (C) \# 26400245.
S.M. would like to thank Yukawa Institute for Theoretical Physics at Kyoto University for warm hospitality.

\end{document}